\documentclass[utf8]{FrontiersinVancouver} 
\usepackage{tabularx}
\usepackage{url,hyperref}
\def\keyFont{\fontsize{8}{11}\helveticabold }
\def\firstAuthorLast{Subedi {et~al.}} %use et al only if is more than 1 author
\def\Authors{Vishal Subedi\,$^{1}$, Shashipraba N. K. Rajakaruna\,$^{2}$, Pratyusha Sarkar\,$^{1}$,  Subhankar Chattoraj\,$^{3}$,  Anjali Khasa\,$^{3}$, Siddhartha Nandy\,$^{1}$, Hamza Farooq\,$^{4}$, Animikh Biswas\,$^{1}$, Sanjay Chaudhuri\,$^{5}$,  Asim K. Dey\,$^{2}$,  Karuna Joshi\,$^{3}$, Christophe Lenglet\,$^{4}$, and Ansu Chatterjee\,$^{1*}$}
\def\Address{
$^{1}$ Department of Mathematics and Statistics, University of Maryland Baltimore County, Baltimore, Maryland, USA \\
$^{2}$ Department of Mathematics and Statistics, Texas Tech University, Lubbock, Texas, USA, \\
$^{3}$ Department of Information Systems, University of Maryland Baltimore County, Baltimore, Maryland, USA \\
$^{4}$ Center for Magnetic Resonance Research, Department of Radiology, University of Minnesota, Minneapolis, Minnesota, USA \\
$^{5}$ Department of Statistics, University of Nebraska Lincoln, Lincoln, Nebraska, USA.
}
\def\corrAuthor{Ansu Chatterjee}

\def\corrEmail{snigchat@umbc.edu}

\begin{document}
\onecolumn
\firstpage{1}

\title[Data-driven neuroscience]{Data-driven techniques for translational neuroscience and personalized neuro-health} 

\author[\firstAuthorLast ]{\Authors} %This field will be automatically populated
\address{} %This field will be automatically populated
\correspondance{} %This field will be automatically populated

\extraAuth{}

\maketitle

\begin{abstract}

\section{}
Neurodegenexrative diseases such as Alzheimer's disease and Parkinson's disease are diagnosed most reliably only after substantial, often irreversible, neuronal loss has already occurred, creating an urgent need for quantitative tools that can detect subtle, early, and individual-specific brain changes from neuroimaging data. This review surveys a broad and rapidly evolving toolkit of data-driven techniques for translational neuroscience and personalized neuro-health, organized around four complementary methodological pillars. We first review classical and Bayesian regression-based approaches to functional MRI (fMRI) analysis, including mass-univariate and spatially informed Bayesian models, and illustrate an empirical-likelihood-based alternative that relaxes strong parametric assumptions; we further describe covariate-assisted principal regression and its Bayesian extension, which link functional connectivity directly to individual-level demographic and clinical covariates, a key ingredient for personalized, digital-twin-style models of brain function. We then turn to methods that exploit the geometric and topological structure of the brain, including discrete Ricci curvature as a biomarker of structural and functional network robustness, topological data analysis and persistent homology for characterizing multiscale organization in structural MRI and brain networks, and knowledge-graph and semantic-web representations that support clinical data interoperability. Next, we review artificial intelligence techniques for neuroimage-based diagnosis and modeling, spanning convolutional and recurrent neural networks, hybrid and multimodal deep architectures, generative adversarial networks and diffusion models for synthetic data generation and disease-progression modeling, and both classical and deep-learning-based approaches to longitudinal change detection. Finally, we review dynamical systems and mechanistic models of neuroimaging data, from single-neuron and neural-mass models to whole-brain network models and disease-progression dynamics, together with the statistical and machine-learning tools used to learn such systems from data. Throughout, we emphasize how these methodologically diverse approaches converge on a common translational goal: personalized, mechanistically grounded, and clinically actionable models of individual brain health, and we close by discussing the principal open statistical, computational, and clinical challenges that remain.

\tiny
 \keyFont{ \section{Keywords:} precision medicine, Bayesian statistics, topological data analysis, network curvature, knowledge graphs, generative AI, dynamical systems, digital twin} 
\end{abstract}

\section{Introduction}
\label{Sec:Introduction:AC}

Neurodegenerative diseases have become increasingly common worldwide, impacting millions of people and involve enormous social and economic consequences \citep{zahra2019global, morris2021neuroscience20, arias2022quantifying}. Two of the most prevalent such diseases are Alzheimer's disease, the leading cause of dementia, and Parkinson's disease, the second most common neurodegenerative disorder; both face considerable difficulty in early diagnosis, precisely when therapeutic interventions may be most effective \citep{rascol2011double, van2023lecanemab, ghosh2024review}. This diagnostic gap has motivated an increasingly rich and interdisciplinary body of quantitative methodology --- spanning classical and Bayesian statistics, differential geometry and topology, artificial intelligence, and dynamical-systems theory --- aimed at extracting earlier, more sensitive, and more individually meaningful signals from neuroimaging data. A recurring theme throughout this literature, and throughout this review, is a shift away from purely inferential descriptions of \emph{where} in the brain a stationary statistical effect is located, and toward mechanistic, personalized, and dynamic characterizations of \emph{how} an individual's brain organization and function evolve, and how they are altered by disease \citep{john2022}.

This review is organized into four main parts, each surveying a distinct but complementary methodological pillar of data-driven, translational neuroscience. Section~\ref{Sec:fMRI_Basics} reviews functional MRI (fMRI) data analysis and dimension-reduction techniques. We begin with the classical mass-univariate regression and statistical parametric mapping (SPM) framework, dating to the characterization of the blood-oxygen-level-dependent (BOLD) contrast mechanism \citep{ogawa1990} and its subsequent formalization as a general linear model fit independently at every voxel \citep{friston1994spm}, together with its Bayesian counterparts, which introduce spatial and hierarchical structure directly into the model \citep{friston2002classical1}. We then illustrate an alternative, empirical-likelihood-based approach to voxel-wise inference \citep{owen2010}, which relaxes the strong parametric assumptions underlying classical and Bayesian regression and constructs a data-driven null distribution directly from background (non-activated) voxels. Section~\ref{Sec:fMRI_Basics} closes with a detailed treatment of covariate-assisted principal (CAP) regression and its Bayesian extension (BCAP) \citep{zhao2021covariate, park2025bcap}, which explicitly link low-dimensional representations of functional connectivity to individual-level covariates such as age and sex --- a capability that is foundational to the construction of personalized, digital-twin-style computational models of brain function.

Section~\ref{Sec:GraphsETC} reviews approaches that draw on graph theory, network science, differential geometry, and topology. We describe the application of discrete Ricci curvature --- both the Ollivier--Ricci \citep{ollivier2007ricci} and Forman--Ricci \citep{forman2003bochner} notions --- as a geometric biomarker of structural and functional network robustness and fragility \citep{farooq2019network}, with demonstrated sensitivity to healthy aging and to conditions including autism spectrum disorder, multiple sclerosis, and attention-deficit/hyperactivity disorder, as well as recent theoretical extensions relating network curvature to community structure and to broader biomedical network science \citep{albert2026analyzing, brooks2024community}. We then review topological data analysis and, in particular, persistent homology \citep{carlsson2009}, which characterizes the multiscale topological organization of structural MRI, resting-state and task fMRI, and structural/functional brain networks, with applications ranging from tumor morphology \citep{crawford2020predicting} to Alzheimer's disease classification \citep{saadat2021topological}, and provide a broader survey of this rapidly growing area \citep{singh2023tda}. Finally, we discuss knowledge-graph and semantic-web representations \citep{antoniou2009owl, pan2009resource}, which support the interoperable exchange of clinical neuroimaging data, models, and usage policies across institutions \citep{chattoraj2024semantically}, an increasingly important requirement for multi-site, translational research.

Section~\ref{Sec:AI_Neuro} reviews artificial intelligence techniques for neuroscience studies, with particular emphasis on their application to neurodegenerative disease. We survey classical deep learning architectures --- convolutional neural networks, recurrent and long short-term memory networks, and hybrid and multimodal deep architectures --- as applied to the diagnosis, staging, and progression modeling of Alzheimer's and Parkinson's disease \citep{chaki2023deep, olaniyan2023new, sharma2023deep, islam2018brain}. We then turn to generative artificial intelligence, including generative adversarial networks, variational autoencoders, and denoising diffusion models, which support synthetic neuroimage generation, data augmentation, and interpretable disease-progression modeling \citep{zong2024brainnetdiff}, and we review both classical, registration-based techniques and modern deep-learning approaches to the detection of longitudinal change in serial medical images \citep{patriarche2004review, hadley2016change, lachinov2023learning}, a task of direct relevance to monitoring disease progression and treatment response.

Section~\ref{Sec:DynSys_Neuro} reviews dynamical systems and mechanistic models of neuroimaging data. In contrast to the largely stationary, localized statistical descriptions of the other sections, this literature explicitly models the brain as an evolving dynamical system, from biophysical single-neuron and neural-mass models through whole-brain network models to explicit models of neurodegenerative disease progression \citep{alexandersen2026network, rollo2023dynamical, raj2012, garbarino2021investigating}, together with the statistical and machine-learning tools --- Kalman filtering, variational Bayesian estimation, and their modern extensions --- used to learn such systems from noisy, high-dimensional neuroimaging data. This section also introduces the emerging paradigm of personalized, generative \emph{digital brain twins} \citep{zimmermann2022, bakr2024synthetic, wang2024virtualbraintwins}, which draws together the individualized, covariate-informed modeling of Section~2, the mechanistic models of Section~~\ref{Sec:DynSys_Neuro}, and the AI-based estimation tools of Section~~\ref{Sec:AI_Neuro} into a single translational framework.

Across these four pillars, a common thread emerges: the growing recognition that classical statistical inference \citep{lindquist2008, poldrack2011handbook}, geometric and topological representation, artificial intelligence, and mechanistic dynamical modeling are not competing alternatives but complementary tools, each addressing a different facet of the shared translational goal of personalized neuro-health.

Owing to the breadth and diversity of the topics covered, this paper is unfortunately notation and acronym-heavy. In Table~\ref{Tab:Acronym} we have provided a list of common acronyms that are used throughout the paper and is commonly used across various related topics. A few other names and acronyms are embedded in the text below as and when they arise: such cases arise  only a small number of times in this paper. 
\begin{table}[h]
    \centering
    \begin{tabular}{|l|l|}
    \hline
         Notation/Acronym &  Definition (expanded form)\\
         \hline
         \hline
         & Neuroscince related terms \\
         \hline
         \hline
         ADHD & attention deficit hyperactivity disorder\\
         AD & Alzheimer's disease \\
         ADNI & Alzheimer's Disease Neuroimaging Initiative \\
         ASD & autism spectrum disorder \\
         dMRI & diffusion magnetic resonance imaging \\
         EEG & electroencephalography \\
         fMRI & functional magnetic resonance imaging \\
         MCI & mild cognitive impairment\\
         MEG & magnetoencephalography\\
         MRI & magnetic resonance imaging \\
         MS & multiple sclerosis \\
         PD & Parkinson's disease \\
         PET & positron emission tomography\\
         \hline
         \hline
         & Data science related terms \\
         \hline
         \hline
         AI & artifical intelligence\\
         CNN & convolutional neural network \\
         DNN & deep neural network \\
         GAN & generative adversarial network\\
         ICA & independent component analysis\\
         LLM & large language model \\
         LSTM & long-short term memory \\
         ML & machine learning\\
         NNMF & non-negative matrix factorization \\
         PCA & principal component analysis\\
         RNN & recurrent neural network \\
         TDA & topological data analysis \\
         VAE & variational autoencoder\\
         \hline
    \end{tabular}
    \caption{Some commonly used acronyms used in this paper. Some additional acronyms, that are less commonly used, are embedded in text as and when they arise.}
    \label{Tab:Acronym}
\end{table}

% #####################################################################
\section{{f}MRI data analysis and dimension reduction techniques}
% #####################################################################
\label{Sec:fMRI_Basics}

Functional magnetic resonance imaging (fMRI) exploits the blood-oxygen-level-dependent (BOLD) contrast mechanism to produce indirect, hemodynamically-mediated measurements of neural activity, and serves as a major quantitative tool in neuroscience studies. Often fMRI studies are conducted when the (human) subject performs a task (like seeing human faces or scrambled images), or when the subject is in a resting state. The data is typically indexed with a four-dimensional tensor (three spatial dimensions plus time), but may also contain information on the details of the task, features associated with the subject like their age, sex and so on, and other important information. As an example, consider the study of \cite{wakeman2015multi}, where multiple subjects go through multiple fMRI scans, and in addition, EEG and MEG scans and scans to obtain brain structural data. Alongside,  information on the visual tasks and subject characteristics are available. For the illustrative examples, below, we use part of this dataset, obtained from the OpenfMRI database (accession number ds000117). The fMRI scans are recorded over a (64, 64, 33)-dimensional grid in three dimensions (such grid cells are called \textit{voxels}), and hence is high-dimensional, as is typical of most fMRI studies. Detailed exposition on neuroimage data and FMRI data in particular may be found in \cite{lindquist2015handbook, lindquist2008, lazar2008, polzehl2019}. 

One of the main approaches towards analyzing fMRI data is based on using linear statistical models, where the same regression model is fit at every voxel, producing whole-brain \textit{statistical parametric maps} (SPMs) of the evidence for experimental effects. While this approach remains the backbone of fMRI data analysis, a Bayesian framework may also be adopted to include detailed information on spatio-temporal dependencies, dependencies between multiple scans on the same subject, multi-subject models. We briefly review both these approaches in this section, before providing insights into an alternative modeling approach using empirical likelihood, and some techniques on dimension reduction of fMRI data.

% #####################################################################
\subsection{fMRI regression and SPM-based techniques and extensions}
% #####################################################################

In a pioneering work,  \cite{ogawa1990} characterized the blood-oxygen-level-dependent (BOLD) contrast mechanism, demonstrating that the differential magnetic susceptibility of oxygenated and deoxygenated hemoglobin could be exploited to produce MR image contrast that varies with local blood oxygenation. Other major developments are provided in \cite{kwong1992, buxton1998}, and \cite{bandettini1993, friston1993connectivity, friston1995eigenimage} developed some of the first systematic processing strategies for the resulting time-course data, including using correlations, PCA, ICA, Fourier analysis methods and so on for detecting task-related signal changes voxel by voxel. These early studies established fMRI as a viable tool for functional brain mapping. 

The next stage of advancement was in the development of standard data preprocessing pipelines and the statistical parametric mapping (SPM) framework \citep{boynton1996, friston1994spm, worsley1995, friston1995analysis, friston1996movement, behzadi2007, polzehl2000, polzehl2006, tabelow2006, polzehl2010, tabelow2009, tabelow2014, yue2010}, and the understanding of the the data preprocessing pipeline that is often needed. Such preprocessing may include slice-timing correction, motion correction, distortion correction, spatial normalization and smoothing, structural adaptive smoothing, temporal filtering and other steps.

The dominant framework for classical fMRI statistical analysis is the mass-univariate linear model.
 Consider data where there are $I$ individuals in the study, who are denoted by $i \in \{ 1, \ldots, I \}$.  Each subject has $J$ fMRI scans while performing a task, and these images are enumerated as $j \in \{ 1, 2, \ldots, J \}$.  Each scan is over a collection of voxels indexed by $(\ell_{x}, \ell_{y}, \ell_{z}) \in  \mathbb{L} = \{ 1, \ldots, L_{x} \} \times  \{ 1, \ldots, L_{y} \} \times \{ 1, \ldots, L_{z} \}$. Define $L = L_{x} L_{y} L_{z}$. To simplify the notation, we first list the voxels in a one-dimensional array.  To that goal, define $\ell = L_{y} L_{z} (\ell_{x} - 1) + L_{z} (\ell_{y} -1)+ \ell_{z}$.  The mapping between $(\ell_{x}, \ell_{y}, \ell_{z}) \in  \mathbb{L}$ and $\ell \in \{ 1, \ldots L \}$ is bijective, and thus it is equivalent to index voxels using $\ell \in \{ 1, \ldots L \}$. For the $i$-th subject and $j$-th scan, in each voxel $\ell \in \{ 1, \ldots, L \}$, the response is a time series $ \{ Y^{(i, j)}_{t,\ell}: \ t = 1, \ldots, T \}$ of length $T$.  For each voxel $\ell$, we collect its response in a $T\times 1$ vector $\mathbf{Y}^{(i, j)}_{\ell}$. Furthermore, for the $i$-th subject and $j$-th scan, there is also a task-related covariate or feature matrix $\mathbf{X}^{(i, j)} \in \mathbb{R}^{T \times P}$, that encapsulates in its rows the task-related features associated with the scan at time $t \in \{ 1, \ldots T \}$. Our goal is to find the voxels  which are activated during the task. The dataset of \cite{wakeman2015multi} can serve as a conceptual illustration for this section. 

We assume a linear regression model for the temporal response at each voxel. Specifically, for the $i$-th subject and $j$-th scan, the response vector for each voxel $\ell \in \{ 1, \ldots, L \}$ is modeled as:
    \begin{align}
        \mathbf{Y}^{(i, j)}_{\ell} = \mathbf{X}^{(i, j)} \beta^{(i, j)}_{\ell} + \epsilon^{(i, j)}_{\ell}, \label{eq:mainModel} 
    \end{align}
    where $\beta^{(i, j)}_{\ell} =\left(\beta^{(i, j)}_{1,\ell},\beta^{(i, j)}_{2,\ell},\ldots,\beta^{(i, j)}_{P,\ell}\right)^{\prime}\in \mathbb{R}^{P}$ is the vector of regression coefficients, and $\epsilon^{(i, j)}_{\ell} \in \mathbb{R}^{T}$ is the vector of noise terms which could be mutually dependent and often modeled with an autoregressive process in each voxel. Least squares estimates are obtained for each voxel. Additional discussions and details may be found in \cite{lazar2008, lindquist2008, monti2011, poline2012} and other places. Some extensions of the above setup are available in \cite{haxby2001, norman2006, friston2011connectivity}.

Statistical inference in this approach may strongly depend on the model assumptions, and multiple comparison corrections \citep{friston1994extent, genovese2002, nichols2002, smith2009, chen2013}. The reliability of these procedures have been critically examined in \cite{eklund2016}.

\subsection{Bayesian Inference for fMRI}

Let $\theta$ denote all the parameters in the above mass-univariate framework. In a Bayesian model, a likelihood function related the observed data $\{ \mathbf{Y}^{(i, j)}_{\ell} \}$ to this parameter as in \eqref{eq:mainModel}, while a prior distribution over the parameter space captures the investigator's subjective belief. Except in the simplest of  settings, the posterior $p({\theta}\mid \{ \mathbf{Y}^{(i, j)}_{\ell} \})$ is analytically intractable at scale. However,   \textit{Markov chain Monte Carlo} (MCMC)-based inference  for the whole-brain, voxel-wise spatiotemporal models is studied in \citep{musgrove2016, bezener2018}, while related discussions on variational Bayes (VB) techniques can be found in \citep{penny2003variational, friston2002dcmestimation, woolrich2012review}. General expositions of the classical-versus-Bayesian contrast, and its specific implications for neuroimaging inference, are given by \cite{friston2002classical1} and \cite{friston2002classical2}.

The most direct Bayesian analogue of the classical mass-univariate GLM replaces the independent, voxel-wise estimation of regression coefficients in \eqref{eq:mainModel} with a joint model where a spatial prior couples coefficients across neighboring voxels, so that estimation at each location borrows statistical strength from its neighbors, while further developments involve more complex spatio-temporal extensions \citep{genovese2000, gossl2001, woolrich2004fully, penny2005, penny2007spatial, penny2003mixtures}. In \cite{flandin2007}, a Bayesian fMRI regression model is developed using sparse spatial basis function priors, which both regularizes the spatial pattern of estimated activation and naturally achieves a form of dimension reduction.

Bayesian hierarchical (multilevel) models and related developments are presented in \citep{woolrich2004multilevel, beckmann2003multilevel, bowman2007, bowman2008, guo2008predicting, degras2014}. Bayesian variable selection methods and related discussions are available in \cite{smith2007, goldsmith2014, zhang2016variableselection}. Joint spatiotemporal Bayesian modeling is studied in \cite{gossl2001, woolrich2004fully, quiros2010, derado2010, sanyal2012, musgrove2016, bezener2018}.  A distinct but related joint-modeling literature addresses regression problems in which an image enters as a predictor or response \citep{reiss2010, zhu2014, kang2018}, where various Bayesian scalar-on-image regression framework are developed. 
Additional discussion on Bayesian modeling of fMRI data may be found in \cite{caffo2010, lindquist2014dynamic, lindquist2013ironic, bowman2008}. Empirical comparisons between classical and Bayesian methods on shared data or shared inferential criteria, rather than developing either paradigm in isolation can be found in \cite{penny2005, woolrich2004fully}. See also \cite{friston2012ironic} and its principled rebuttal \cite{lindquist2013ironic}.

The popular neuroimaging software suites SPM, FSL, and AFNI that implement the classical and Bayesian regression methods are discussed in \cite{penny2007spm, woolrich2001, poldrack2011handbook}. In \texttt{R}, related packages include \texttt{fmri, oro.nifti, oro.dicom, fslr, ANTsR, neuRosim, dti}, discussed in \cite{tabelow2011, whitcher2011,  avants2011, welvaert2011, muschelli2015}. Similarly, in \texttt{Python}, available packages include 
\texttt{NiBabel, Nipype, fMRIPrep, Nilearn, PyMVPA, BIDS}, discussed in \cite{gorgolewski2011, esteban2019, abraham2014, hanke2009, gorgolewski2016}.

\subsection{An illustrative example of empirical likelihood inference in fMRI data}
\label{Sec:SC:EL}

We now illustrate how modern data science approaches can provide considerable versatility and robustness in understanding and analyzing neuroimage data, enhancing the kind of models discussed above. 
Suppose that in the context of the data of \cite{wakeman2015multi}, based on the model in \eqref{eq:mainModel} and for a given $p=p_0$, we want to test the null hypothesis:
\begin{equation}\label{eq:mainNull}
    H_0 : \beta^{(i,j)}_{p_0\ell} = 0 \quad \text{for each voxel, i.e., } \ell=1,2,\ldots,L.
\end{equation}
Using the concept of empirical likelihood (EL) \citep{owen2010}, we define the test statistic as:
\begin{equation}\label{eq:T}
\mathcal{T}^{(i,j)}_{p_0\ell} = \max_{w\in\mathcal{W}} \left\{ -2 \sum^T_{t=1} \log(T w_t) \right\},
\end{equation}
where the weight vector $w=(w_1,w_2,\ldots,w_T)^{\prime}$ is maximized over the constraint set $\mathcal{W}$, defined by: 
\begin{equation}\label{eq:W}
    \mathcal{W} = \bigcup_{\tilde{\beta}^{(i,j)}_{\ell}\in\mathbb{R}^{p-1}} \left\{ w : \left(\mathbf{X}^{(i, j)}\right)^{\prime} D_{w} \left(\mathbf{Y}^{(i, j)}_{\ell} - \tilde{\mathbf{X}}^{(i, j)} \tilde{\beta}^{(i, j)}_{\ell}\right) = 0 \right\} \cap \Delta_{T-1},
\end{equation}
where $\tilde{\beta}^{(i, j)}_{\ell}\in\mathbb{R}^{p-1}$ is the vector of regression coefficients excluding the $p_0$-th entry, $\tilde{\mathbf{X}}^{(i, j)}$ is the $T\times(p-1)$ submatrix of $\mathbf{X}^{(i, j)}$ with its $p_0$-th column removed, and $D_{w}$ is the $T\times T$ diagonal matrix with $w$ along its diagonal.  
A key part of the constraints defining the set $\mathcal{W}$ consists of the weighted score functions corresponding to the model in \eqref{eq:mainModel}. We additionally require that the weighted residuals be orthogonal to all $p$ columns of $\mathbf{X}^{(i, j)}$ \citep{chenWangWuLi2022}.
The test statistic in \eqref{eq:T} is computed by maximizing the objective function over all values of $w\in\mathcal{W}$. Since $\mathcal{W}$ is defined as a union over all $\tilde{\beta}^{(i, j)}_{\ell}\in\mathbb{R}^{p-1}$, computing $\mathcal{T}^{(i,j)}_{p_0\ell}$ also requires maximizing the objective in \eqref{eq:T} with respect to $\tilde{\beta}^{(i, j)}_{\ell}$.

If the entries of $\mathbf{Y}^{(i, j)}_{\ell}$ were independent, under standard regularity conditions, $\mathcal{T}^{(i,j)}_{p_0\ell}$ would asymptotically follow a $\chi^2_{1}$ distribution as $T\rightarrow\infty$. The p-value $\pi^{(i,j)}_{p_0\ell}$ for each $\ell=1,2,\ldots,L$ could then be estimated from this asymptotic distribution. However, due to the temporal/spatial dependence inherent in the fMRI response, the true asymptotic distribution may differ. Therefore, we determine the null distribution of the test statistics empirically, directly from the data. Let $\mathcal{B}$ denote the subset of voxels within the brain region, assumed to be known from the masking information. Conversely, any voxel in the complement set $\mathcal{B}^C$ resides in the background void and exhibits no task-related activity. Consequently, these background voxels naturally satisfy the null hypothesis in \eqref{eq:mainNull}. 

The null distribution of the test statistic is empirically constructed from the observed values of $\mathcal{T}^{(i,j)}_{p_0\ell}$ for $\ell\in\mathcal{B}^C$. For any target voxel $\ell^{\star}\in\mathcal{B}$, we estimate its raw p-value, $\pi^{(i,j)}_{p_0\ell^{\star}}$, as the proportion of observed background test statistics that are greater than or equal to $\mathcal{T}^{(i,j)}_{p_0\ell^{\star}}$. Because an empirical p-value cannot strictly equal zero by convention, we employ a standard pseudo-count correction to estimate $\pi^{(i,j)}_{p_0\ell^{\star}}$ as:
\begin{equation}\label{eq:rawPval}
  \hat{\pi}^{(i,j)}_{p_0\ell^{\star}} = \frac{ \left| \left\{ \ell \in \mathcal{B}^C : \mathcal{T}^{(i,j)}_{p_0\ell^{\star}} \le \mathcal{T}^{(i,j)}_{p_0\ell} \right\} \right| + 1}{\left| \mathcal{B}^C \right| + 1}.
\end{equation}
By construction, for all $\ell^{\star}\in\mathcal{B}$, the estimated p-values are bounded below such that $\hat{\pi}^{(i,j)}_{p_0\ell^{\star}} \ge \left( \left| \mathcal{B}^C \right| + 1 \right)^{-1}$.
Once $\hat{\pi}^{(i,j)}_{p_0\ell^{\star}}$ is obtained for all $\ell^{\star}\in\mathcal{B}$, the p-values are adjusted using the Benjamini-Hochberg method \citep{BenjaminiHochberg1995} to control the false discovery rate (FDR). We denote these adjusted p-values by $\tilde{\pi}^{(i,j)}_{p_0\ell^{\star}}$ for $\ell^{\star}\in\mathcal{B}$. Finally, voxel-wise activation maps are generated based on these FDR-adjusted p-values.
\begin{figure}[ht]
\begin{center}
\begin{tabular}{ccc}
    \includegraphics[width=0.3\textwidth]{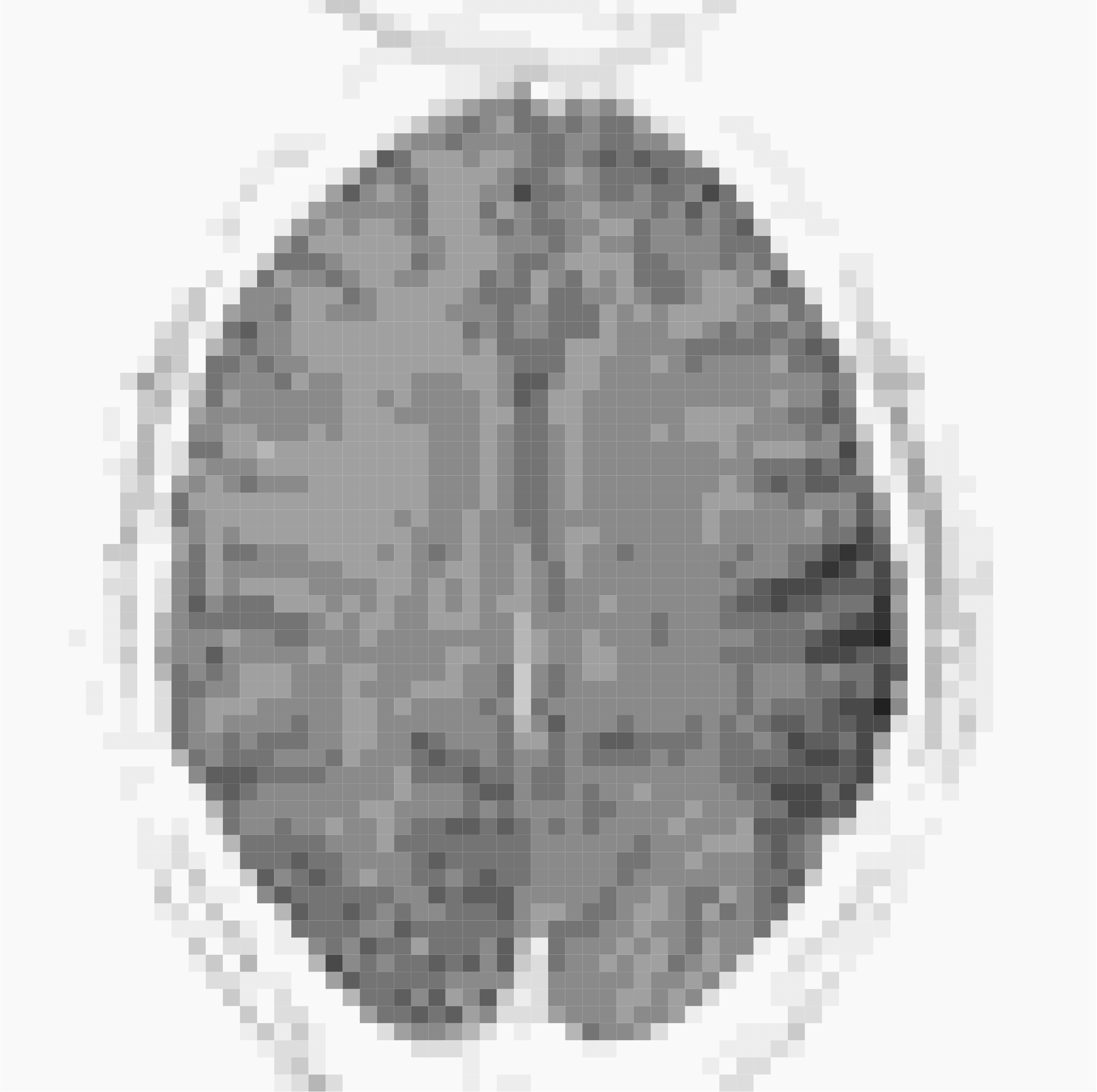} 
    \includegraphics[width=0.3\textwidth]{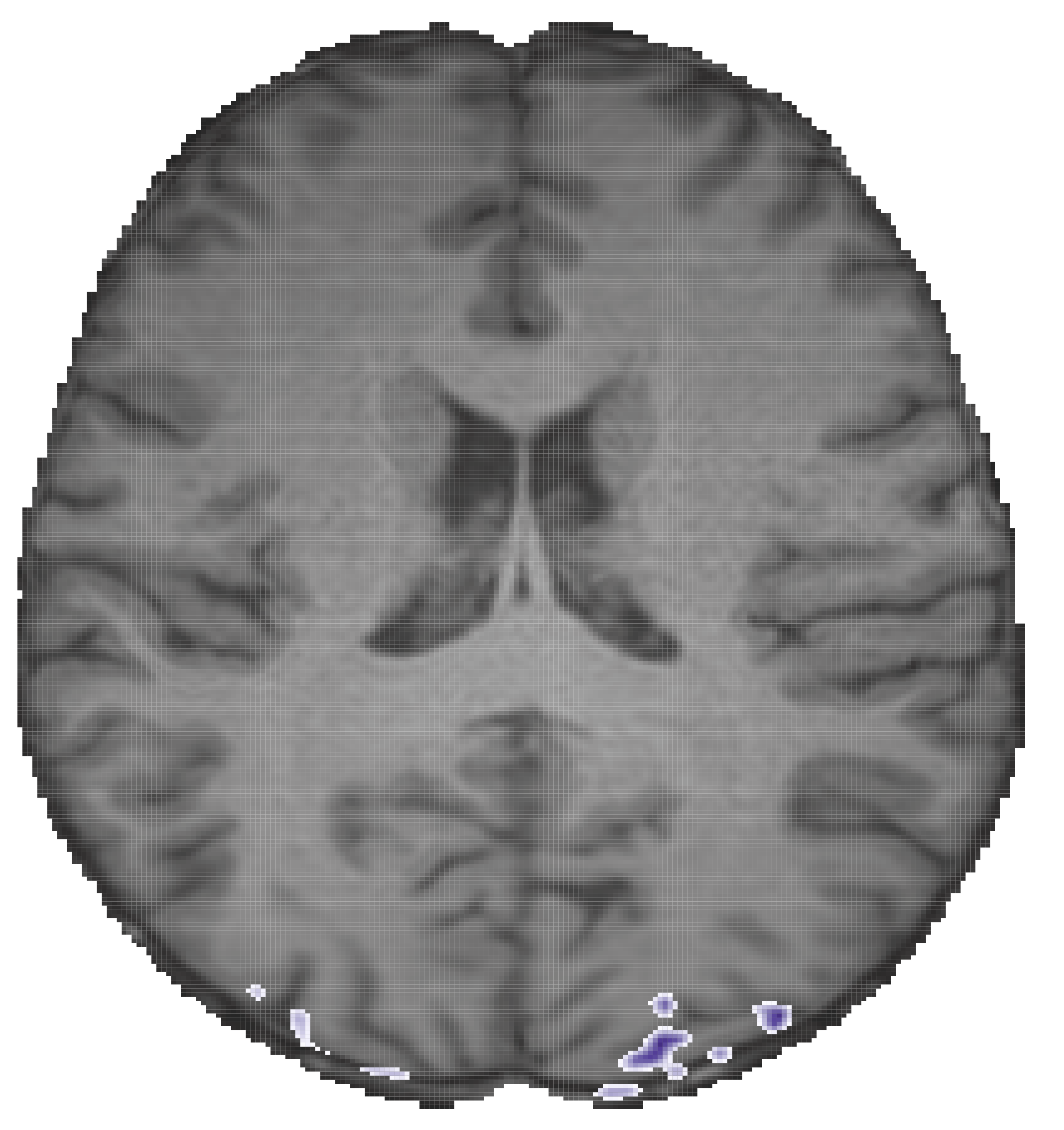} & 
    \includegraphics[width=0.3\textwidth]{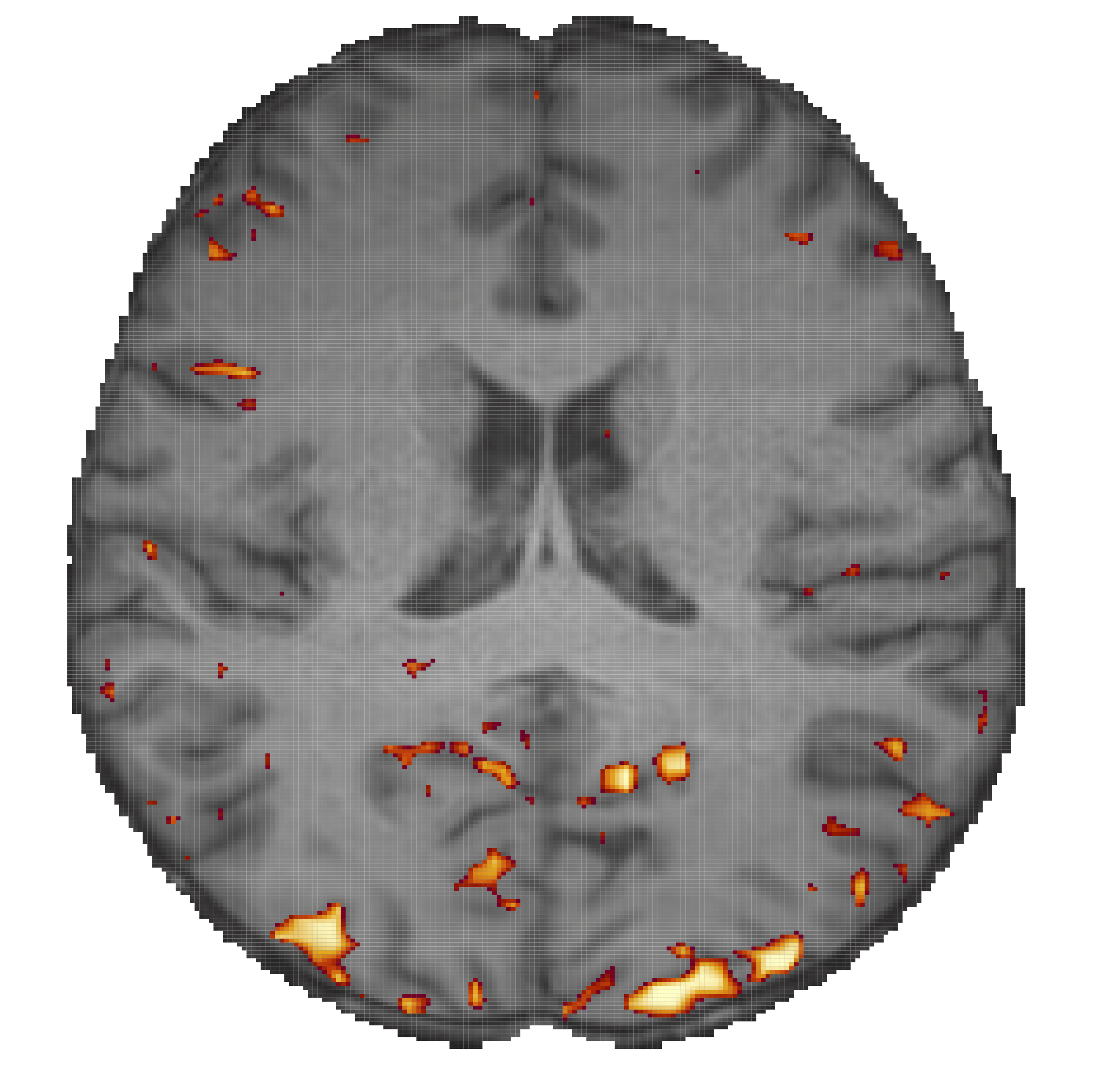}
\end{tabular}
\end{center}
  \caption{An illustrative example of empirical likelihood-based inference in fMRI data. The left panel shows the masked region. The middle panel shows active voxels obtained using linear regression model in an axial slice, and the right panel is the its equivalent obtained using EL techniques.}
  \label{Fig:ELFMRI}
\end{figure}

As an illustration, this  procedure is applied to one scan of the first subject from dataset of \cite{wakeman2015multi}. 
An axial slice of the voxel-wise mean of the $210$ raw observations is presented in the left panel of Figure~\ref{Fig:ELFMRI}.  Heightened activity is represented by the darker shades of gray.  The masked region $\mathcal{B}$ is clearly identifiable. The other  graphics in this figures depict active voxels during visual tasks.  Note that there is a difference in active voxels corresponding to different visual tasks, and the EL-technique provides more robust alternative to parametric statistical frameworks that come with strong assumptions about the data and model.

\subsection{Covariate assisted principal regression for brain functional connectivity}
\label{Sec:SN:CAP}

The development of digital twins in healthcare necessitates the creation of highly accurate, personalized \textit{in silico} models of human physiology. In computational neuroscience, a digital twin of the human brain aims to simulate and predict functional dynamics and connectivity patterns. However, brain organization is highly chaotic; to be clinically and scientifically valid, these computational models must explicitly account for how individual-level biological and demographic traits—such as age and sex affect brain activity. 
Since fMRI data is massively high-dimensional, encompassing hundreds of thousands of voxels, a standard foundational step in network modeling involves brain parcellation: grouping voxels into a smaller set of $K$ distinct \textit{Regions of Interest} (ROIs). By reducing the spatial dimensionality, we can estimate functional connectivity as the covariance or correlation between these regional time series. 
Discussions related to dimension reduction in fMRI data can be found in 
 \cite{mckeown1998, beckmann2004, beckmann2005resting, calhoun2001, guo2008predicting, guo2011, viviani2005, flandin2007, zhang2015review}.

While traditional unsupervised methods like PCA or ICA are frequently used to extract latent functional networks from these massive covariance matrices, they evaluate connectivity avoiding subject-level covariates. Consequently, the extracted components may capture broad population variance or scanner noise rather than the specific functional circuits that vary systematically with demographic or clinical traits, and thus have limited applicability in personalized medicine. To address this limitation, \textit{Covariate-Assisted Principal} (CAP) regression, and its Bayesian extension (BCAP) have emerged as powerful frameworks. 
Despite the statistical robustness of covariate-assisted models, their efficacy is inherently bound by the quality and nature of the input features.

The CAP regression framework, introduced by \cite{zhao2021covariate}, integrates PCA with generalized linear modeling to study multiple covariance matrix outcomes. 
Let $y_{it}\in \mathbb{R}^{p}$, for $i=1,\ldots,n$ and $t=1,\ldots,T_i$, denote independent and identically distributed samples from a multivariate normal distribution with mean zero and covariance matrix $\Sigma_i$. In our context, $y_{it}$ represents the fMRI BOLD measurements across $p$ ROIs. The CAP model assumes the existence of a rotation vector $\gamma \in \mathbb{R}^{p}$ such that the scalar projection $z_{it} = \gamma^{\top} y_{it}$ satisfies the generalized linear model:
    $\log \mathbb{V}[z_{it}] = \log \gamma^{\top}\Sigma_i \gamma = x^\top_{i}\beta$,
where $x_i \in \mathbb{R}^{q}$ is the subject-specific covariate vector and $\beta \in \mathbb{R}^{q}$ captures the covariate effects. The log-variance link ensures positivity of the variance and stabilizes it, effectively creating a multivariate extension of classical variance regression \cite{harvey1976estimating}.
The parameters $(\beta, \gamma)$ are estimated by minimizing the negative log-likelihood of the projected data:
\begin{align}
\min_{\beta,\gamma} \quad  \frac{1}{2}\sum_{i=1}^{n} T_i \left\{x_i^\top \beta + \gamma^\top \frac{\hat{\Sigma}_i}{\exp(x_i^\top \beta)} \gamma\right\},  \ \ 
\mbox{subject to} \quad  \gamma^\top H\gamma = 1,
\label{eq:CAP_Zhao_Optimizer_Single}
\end{align}
where $\hat{\Sigma}_i$ is the sample covariance matrix and $H$ is typically the weighted average sample covariance matrix $\bar{\Sigma}$. This optimization can be sequentially extended to identify multiple, orthogonal covariate-associated directions, using metrics such as the deviation-from-diagonality (DfD) criterion \cite{flury1986algorithm} to determine the effective number of principal directions.

Despite its conceptual elegance, the optimization problem in equation~\eqref{eq:CAP_Zhao_Optimizer_Single} is inherently Euclidean—it treats covariance matrices as unconstrained elements in $\mathbb{R}^{p \times p}$. However, the space of symmetric positive definite (SPD) matrices, denoted $\mathrm{Sym}^+_p$, is a Riemannian manifold with non-Euclidean geometry. 
To address this geometric constraint, \cite{park2025bcap} developed a Bayesian counterpart (BCAP). Unlike the original formulation which sequentially identifies single 1D projections, BCAP simultaneously extracts a lower-dimensional subspace whose covariance heterogeneity is explicitly linked to the covariates $x_i$. This is represented by an orthogonal projection matrix $\Gamma \in \mathbb{R}^{p \times d}$, satisfying $\Gamma^\top \Gamma = I_d$ with $d \ll p$.
The BCAP framework proposes a generative latent factor representation of the observed fMRI signals:
$y_{it} = \Gamma \Psi^{1/2}_i s_{it} + L_i \epsilon_{it}$,
where the latent factors are standard normal $s_{it} \sim \mathcal{N}(0,I_d)$, and the noise is $\epsilon_{it} \sim \mathcal{N}(0,I_{p-d})$. 
Crucially, the covariate-driven variance is captured by the diagonal matrix $\Psi^{1/2}_i = \exp\left(\mathrm{diag}\left((Bx_i + z_i)/2\right)\right)$, which depends on fixed covariate effects ($B$) and subject-specific random effects ($z_i \sim \mathcal{N}(0,\Omega)$) to account for additional heteroskedasticity. 
The above provides a highly interpretable decomposition of the functional connectivity structure. The matrix $\Gamma$ identifies the principal directions of covariance (PDCs) systematically modulated by age, sex, and other demographics ($x_i$). Meanwhile, $L_i \in \mathbb{R}^{p \times (p-d)}$, constrained to be orthogonal to $\Gamma$, spans the residual directions where neural variation is independent of the studied covariates, effectively separating covariate-driven structured networks from background physiological noise.

In BCAP model, the core model parameters $\Theta = \{ \Gamma, \beta, \beta_0, \Omega \}$ are assigned domain-appropriate prior distributions, and require careful modeling. Note that the projection matrix $\Gamma \in \mathbb{R}^{p \times d}$ is constrained to the Stiefel manifold $\mathcal{V}_{d,p}$. To specify a matrix von Mises-Fisher prior over this non-Euclidean space within the HMC framework, we can employ a parameter expansion strategy. We may sample an unconstrained matrix $U \in \mathbb{R}^{p \times d}$ with independent standard normal entries, $U_{ij} \sim \mathcal{N}(0, 1)$, and then project $U$ onto the Stiefel manifold using the symmetric polar decomposition $\Gamma = U(U^\top U)^{-1/2}$.
This deterministic transformation induces a uniform prior distribution over the Stiefel manifold, which corresponds to an isotropic matrix von Mises-Fisher distribution. For the population-level regression coefficients $\beta$ (capturing the covariate effects) and the intercept $\beta_0$, weakly informative Gaussian priors may be used: $\beta \sim \mathcal{N}(0, \sigma^2_\beta), \quad \beta_0 \sim \mathcal{N}(0, \sigma^2_{\beta_0})$
To model the subject-specific random effects, a non-centered parameterization can be used to improve sampling efficiency. The latent random effects were drawn as $z \sim \mathcal{N}(0, 1)$. The random effect covariance matrix $\Omega$ was decomposed into a diagonal scale matrix and a correlation matrix. The standard deviations were assigned a Half-Cauchy prior, $\Omega_{sd} \sim \text{Cauchy}^+(0, 1)$. The correlation structure was modeled using a Cholesky factor $L_\Omega$, which received an LKJ correlation prior controlled by the shape parameter $\eta$.
Finally, to stabilize the log-variance linear predictor and prevent numerical degeneracy during sampling, a small ridge constant (0.01) was added to the exponentiated variance. Prior to projection, the regional fMRI BOLD signals should be spatially whitened using $\bar{S}^{-1/2}$, the inverse square root of the marginal sample covariance matrix, ensuring that the latent factor extraction was properly scaled.

\subsubsection{An illustrative example}

\begin{figure}[h]
\centering
\begin{tabular}{ccc}
    \includegraphics[width=0.26\textwidth]{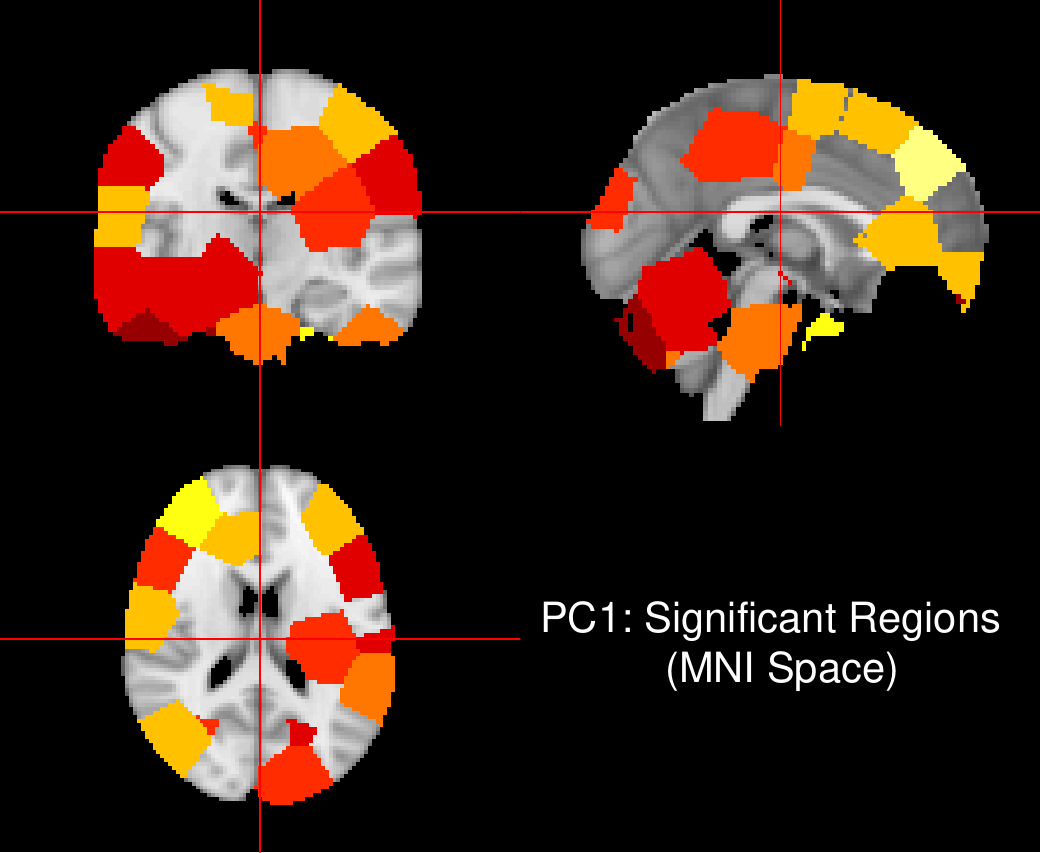} & \includegraphics[width=0.32\textwidth]{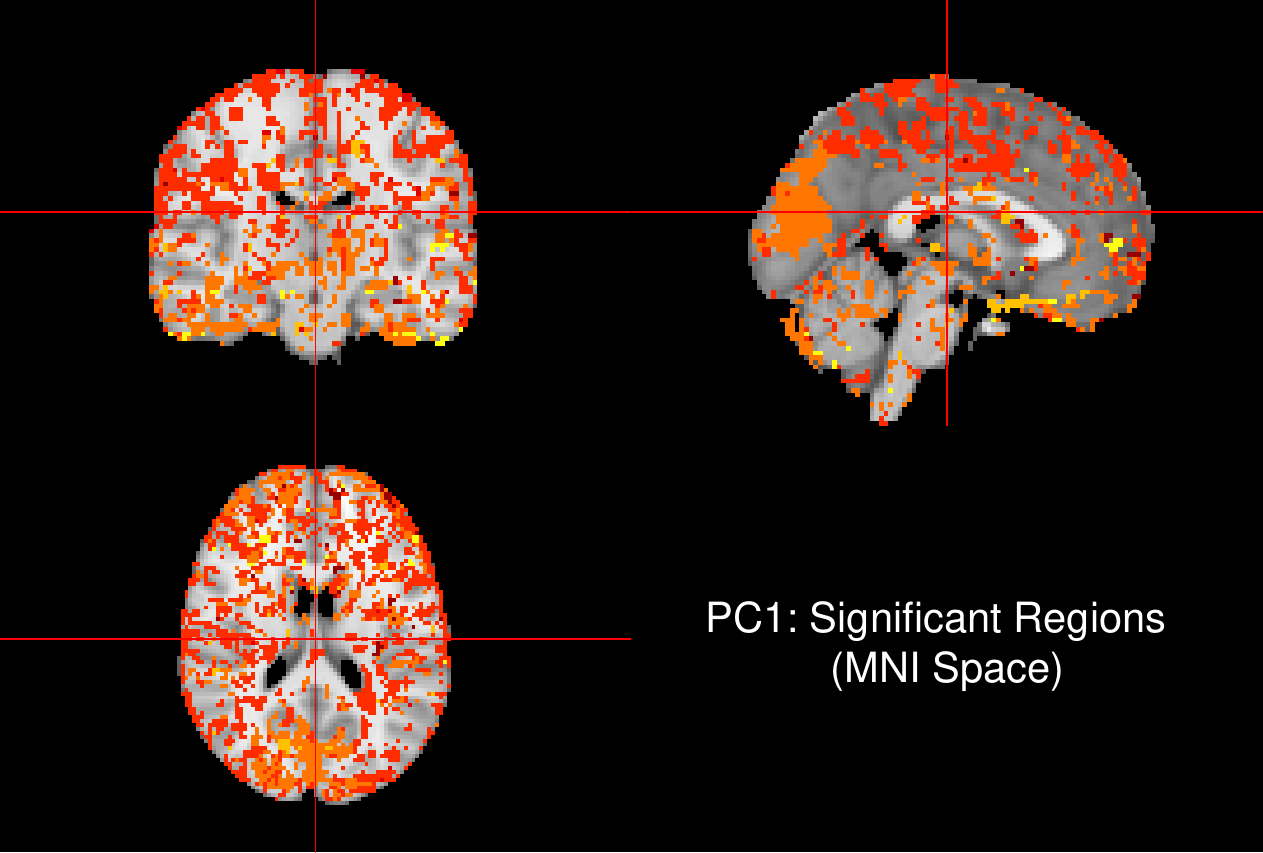} &
    \includegraphics[width=0.26\textwidth]{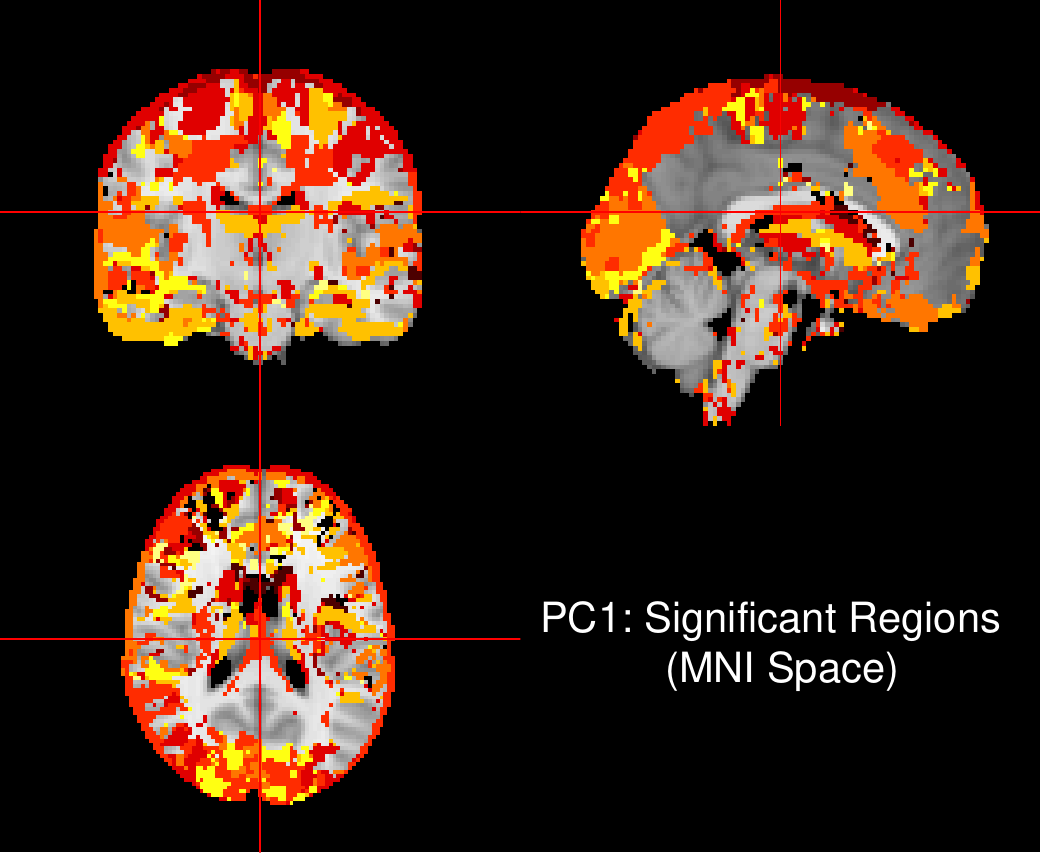}  
\end{tabular}
  \caption{Visualizations of the brain regions with absolute loading values greater than 0.05 for the first Bayesian covariate-assisted principal components, obtained using SSGP, GICA and SCFP parcellation techniques. }
  \label{Fig:PC1_BCAP}
\end{figure}

To ensure spatial correspondence across all subjects ($n=16$) of the data from \cite{wakeman2015multi}, we utilized the standard MNI152 ($2mm$) brain template \citep{jenkinson2001global,jenkinson2002improved}. All subsequent analyses were restricted to voxels falling within the standard binary MNI brain mask $\mathcal{M}$, containing $V$ voxels. We considered Age, Sex ((Male=0, Female=1) and Age-Sex interaction were used as demographic features. To understand the effect of parcellation in subsequent statistical inference and downstream data driven tasks, we partitioned the brain into $p$ distinct parcels using three approaches: ($i$) Sparse Spectral Geometric Parcellation (SSGP, which induces a geometric constraint only)\citep{shi2000normalized,ng2001spectral,von2007tutorial}, ($ii$) Group-Level Spatial ICA (GICA, which captures functional independence) \citep{beckmann2005resting,beckmann2004,calhoun2001} and ($iii$) Spatially-Constrained Functional Parcellation (SCFP, which is a hybrid technique that captures both structure and function) \citep{schaefer2018local,craddock2012whole}.

We mapped the estimated region-specific loadings ($\hat{\Gamma}$) back into the standard 3D MNI coordinate space. The significant regions highlighted in Figure~\ref{Fig:PC1_BCAP} are sensitive to the threshold of the principal component loadings used to create the brain map:  more extensive search to achieve an optimal choice. SSGP yielded no significant demographic association, and posterior credible intervals for Age, Sex, and their interaction included zero. However, both GICA and SCFP obtained that the functional connectivity  differs significantly by sex, with further modulation by age. We illustrate this 
using the first principal components in Figure~\ref{Fig:PC1_BCAP}. For SSGP, despite the lack of demographic modulation, the latent components  still revealed robust, spatially distributed network structures. For GICA and SCFP,  the first principal component represented a distributed network sensitive to demographic factors.  This illustrates that different dimension reduction techniques may obtain different insights from data, and further studies are needed.

\section{Approaches using graphs, networks, topology, geometry}
\label{Sec:GraphsETC}

\subsection{Network curvature in structural and functional brain connectomics}
\label{Sec:HF:NetworkCurvature}

Traditional graph-theoretical measures including degree, clustering coefficient, characteristic path length, modularity, and centrality, have substantially advanced our understanding of brain networks organization. However, these measures provide limited insight into the intrinsic geometry governing information flow, robustness, and resilience. This  limitation has motivated the introduction of geometric concepts from differential geometry into network neuroscience. Among these, \textit{discrete Ricci curvature} has emerged as a powerful framework for characterizing the local geometric organization of complex networks. By quantifying how neighboring regions interact through the transport of probability mass, Ricci curvature provides a measure of local robustness and network fragility that complements conventional graph metrics. More broadly, network geometry has become an active area of complex systems research, extending geometric concepts to understand robustness, community organization, and information flow across diverse biological and technological networks \cite{boguna2021network}. 

\subsubsection{Architectural robustness of structural brain networks}
Structural connectomics maps the physical white matter pathways linking cortical and subcortical regions using dMRI data. Applying network geometry to structural networks allows researchers to treat connectivity as a continuous geometric landscape, where curvature explains architectural stability and physical vulnerability.

The first application of discrete Ricci curvature to dMRI-based structural brain networks was presented by Farooq et al. \cite{farooq2019network}. Using Ollivier-Ricci curvature \cite{ollivier2007ricci,ollivier2009ricci}, the authors proposed a geometric interpretation of structural connectivity in which positively curved regions represent robust network organization, whereas negatively curved regions indicate structurally fragile areas, that are more vulnerable to disruption. Unlike conventional graph measures, curvature simultaneously incorporates local connectivity and neighborhood organization, enabling the identification of brain regions that play critical roles in maintaining global network integrity. The study further demonstrated that curvature-based measures are sensitive to age-related changes and alterations associated with ASD, while remaining consistent with established neuroimaging findings. This work established Ollivier-Ricci curvature as a novel biomarker for assessing the robustness of structural connectomes.

A subsequent study extended this framework to analyze longitudinal data acquired before and after autologous umbilical cord blood infusion in children with ASD \cite{simhal2020measuring}. The analysis demonstrated that curvature detected both local and global changes in structural robustness associated with treatment, including changes that were not identified by traditional graph-theoretical analyses. These findings highlighted the increased sensitivity of geometric descriptors for monitoring subtle structural network changes and reinforced the potential of curvature as a powerful biomarker for longitudinal neuroimaging studies. Further expanding clinical applications, a subsequent work \cite{farooq2020robustness} utilized Ollivier-Ricci curvature to evaluate cognitive impairment in MS. Both local and global curvature metrics successfully differentiated cognitively impaired patients from non-impaired individuals and correlated significantly with overall cognitive decline indices. This work demonstrated that network curvature provides superior insights into structural fragility compared to conventional graph-theoretical metrics in neuro-inflammatory disorders.

\subsubsection{Applications to functional brain networks}

In parallel with developments in structural connectomics, discrete Ricci curvature has also been applied to resting-state and dynamic fMRI networks, providing a geometric framework for characterizing the organization of functional interactions. In particular, Forman-Ricci curvature \cite{forman2003bochner} has been used to identify network anomalies associated with ADHD \cite{chatterjee2021detecting}. The corresponding analysis demonstrated that Forman-Ricci and Ollivier-Ricci curvatures capture complementary geometric properties of functional brain networks and should therefore be regarded as distinct rather than interchangeable measures. Moreover, curvature-based analysis revealed clinically relevant alterations that were not detected using conventional edge-weight analyses, highlighting the ability of geometric metrics to characterize higher-order organizational features of functional connectivity. Applying discrete Ollivier-Ricci \cite{ollivier2007ricci,ollivier2009ricci} and Forman-Ricci curvature \cite{forman2003bochner} to the ABIDE (Autism Brain Imaging Data Exchange) dataset revealed widespread, region-specific functional connectivity alterations in ASD \cite{elumalai2022graph}. Forman-Ricci abnormalities were concentrated within the default mode, somatomotor, and ventral attention networks. Crucially, these affected regions correspond to cognitive domains impaired in ASD and significantly overlap with effective non-invasive brain stimulation targets, underscoring the clinical utility of geometric metrics.

\subsubsection{Characterizing Healthy Aging and Community Detection}

Beyond neuro-developmental disorders, geometric curvature characterizes healthy aging by capturing alterations in resting-state functional connectivity \cite{yadav2023discrete}. Age-related changes in Ollivier-Ricci \cite{ollivier2007ricci,ollivier2009ricci} and Forman-Ricci curvatures \cite{forman2003bochner} occur predominantly in brain regions associated with movement, somatosensory function, and affective processing, with regional curvature values directly correlating with behavioral measures \cite{yadav2023discrete}. This demonstrates that geometric descriptors are sufficiently sensitive to track functional network reorganization during normal aging, supporting their utility across both healthy and pathological brain states.

Additionally, curvature analysis has expanded from identifying localized pathology to characterizing macroscale network organization. Evaluating Ricci flow \cite{hamilton1982three,ollivier2007ricci} based geometric community detection against modularity maximization and Bayesian stochastic block models (SBM) across large-scale fMRI datasets showed that community detection accuracy and robustness depend strongly on underlying network topology \cite{brooks2024community}. This establishes Ricci flow \cite{hamilton1982three,ollivier2007ricci} as an important geometric framework for multiscale network organization, using curvature to study the evolution and overarching structure of brain networks.

The growing adoption of curvature-based methods has also motivated comprehensive evaluations of the various notions of network curvature and their applications across biological systems. A comprehensive evaluation of network curvature notions, including Gromov-hyperbolic curvature \cite{gromov1987hyperbolic}, Ollivier-Ricci curvature \cite{ollivier2007ricci,ollivier2009ricci}, Forman-Ricci curvature \cite{forman2003bochner}, and curvature flow methods \cite{hamilton1982three,ollivier2007ricci}, has highlighted the complementary nature of these geometric frameworks across biological, biomedical, and brain networks \cite{albert2026analyzing}. Collectively, these distinct notions of curvature provide a versatile mathematical framework for characterizing network robustness, community organization, and the dynamical behavior of complex systems \cite{albert2026analyzing}.

\subsubsection{Theoretical Extensions}

Recent theoretical developments have further expanded the scope of network geometry by proposing a unified mathematical framework for describing structural organization, functional dynamics, and disease-related alterations across multiple biological scales. Within this context, curvature-based analysis has been extended beyond brain networks to biomedical knowledge graphs, disease comorbidity networks, and functional brain networks through a geometric phase-transition framework \cite{agourakis2026ollivier}. 
The role of curvature-based methods is rapidly expanding in neuroscience and highlight the potential of network geometry as a unifying mathematical language for studying brain organization across multiple spatial and functional scales. 

\subsubsection{Implementation and Computational Accessibility}

The broader adoption of geometric methods in network neuroscience has been supported by the development of open-source computational frameworks for network construction, analysis, and visualization \citep{centeno2021python,centeno2022hands}. These Python-based tutorials provide practical workflows for analyzing resting-state fMRI data using conventional graph-theoretical measures together with topological data analysis techniques, including tools for the three-dimensional visualization of higher-order interactions projected onto brain atlases. Although these frameworks primarily focus on graph theory and topological data analysis rather than discrete Ricci curvature, they establish scalable computational pipelines that can be readily extended to incorporate curvature-based analyses, thereby facilitating the broader application of geometric approaches in network neuroscience.

\subsection{Topological data analysis (TDA) and related developments in neuroscience}
\label{Sec:AD:TDA}

\subsubsection{Preliminaries on TDA}
\label{sec:TDA_1}

In recent years, the growing complexity and dimensionality of neuroscience data have challenged the capabilities of traditional analytical approaches, including statistical modeling, signal processing, and graph‑theoretical methods. TDA, an emerging framework at the intersection of applied mathematics, statistical inference, and ML, provides a complementary set of tools for characterizing the multiscale organization of neural activity, connectivity, and brain structure.  At the core of this framework is \textit{persistent homology}, which identifies and measures topological signatures such as connected components, loops, voids, and higher‑dimensional cavities. These persistent features offer insight into latent geometric and organizational patterns in neural data that may be obscured when using standard metrics.

Neuroscience data can arise in a variety of forms, including point clouds, functions, or images. To analyze such data topologically, TDA first transforms data into a combinatorial structure known as a \emph{simplicial complex}, which encodes relationships between data points using vertices, edges, triangles, and their higher-dimensional counterparts. 
A $k$-simplex is a basic building block made from $k + 1$ data points. For example, a 0-simplex is a single point, a 1-simplex is a line connecting two points, a 2-simplex is a filled-in triangle formed by three points, and a 3-simplex is a solid tetrahedron formed by four points. One widely used simplicial complex is the \emph{Vietoris–Rips complex}, where a simplex is included if all of its vertices are pairwise within a given distance $\epsilon$. By increasing $\epsilon$, we obtain a nested sequence of simplicial complexes called a \emph{filtration}. This filtration captures how topological features such as connected components, loops, and voids appear and disappear across scales. These topological features are formally described using \emph{homology groups}, which classify the number and type of holes in different dimensions. The rank of the homology groups, known as the \emph{Betti number} $\beta_p$, quantifies the number of $p$-dimensional holes; for example, $\beta_0$ gives the number of connected components, $\beta_1$ gives the number of loops, and so on
~\cite{CohenSteiner2007,Otter_et_al2017,Mileyko_2011,Samantha_2021, carlsson2009,
wasserman2018topological,edelsbrunner2010computational}.

To extract topological features from image data we often use \emph{cubical filtration}, which is well-suited for data represented on a regular grid. A cubical complex is built from elementary cubes (pixels in 2D or voxels in 3D), arranged in a structured lattice. 
Given a grayscale image represented as a function $f: \mathbb{Z}^n \rightarrow \mathbb{R}$, where each voxel is assigned an intensity value, a \emph{sublevel set filtration} is constructed by thresholding this function. Specifically, for each threshold $\alpha \in \mathbb{R}$, the corresponding cubical complex is defined as
$\mathcal{K}_\alpha = \{ x \in \mathbb{Z}^n : f(x) \le \alpha \},$
with the family $\{\mathcal{K}_\alpha\}_{\alpha \in \mathbb{R}}$ forming a nested sequence of cubical complexes indexed by the filtration parameter. As $\alpha$ increases, new cubes enter the complex and existing structures merge or fill in, inducing changes in the topology of $\mathcal{K}_\alpha$ across scales~\citep{kaczynski2006computational,franccois2024train}. 

\begin{figure*}[!ht]
\centering
\includegraphics[width=0.85\textwidth]{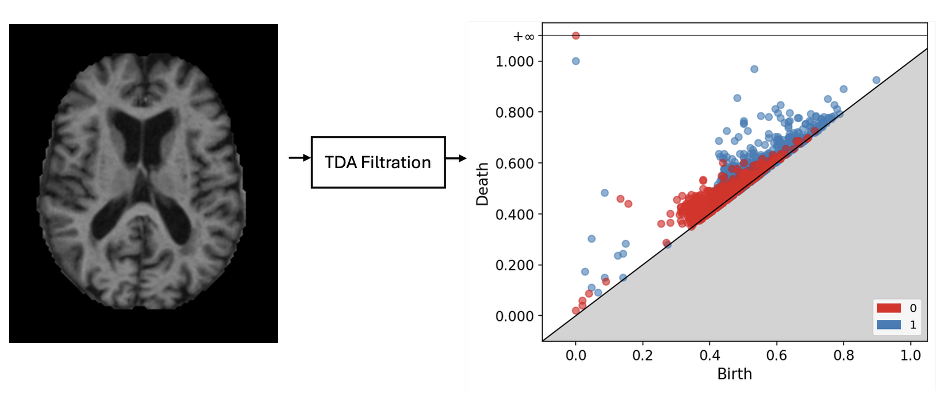}
\caption{Illustration of the TDA steps for an MRI image.}
\label{Fig:TDA:PD}
\end{figure*}

\emph{Persistent homology} tracks these features across the filtration (e.g., Vietoris–Rips filtration and cubical filtration), recording when each feature is born and when it dies. A standard way to represent this information is through a \emph{persistence diagram}, where each topological feature is represented as a point $(\epsilon^b_i, \epsilon^d_j)$ in the plane, corresponding to its birth and death scales. Formally, the persistence diagram is defined as:
\begin{equation}\label{eq:PD}
\mathcal{D} = \{ (\epsilon^b_i, \epsilon^d_j) \in \mathbb{R}^2 \mid \epsilon^b_i < \epsilon^d_j \}.
\end{equation}
Each point in $\mathcal{D}$ thus summarizes a persistent topological feature, providing a compact, multiscale representation of the  structure of the data~\citep{Berwald_2018,panaretos2019statistical,Dey_2023}. Figure~\ref{Fig:TDA:PD} gives a schematic illustration of this construction for a structural MRI image. 
These events are summarized in the persistence diagram  (right panel of Figure~\ref{Fig:TDA:PD}), where red points represent 0-dimensional connected components and blue points represent 1-dimensional features, each encoded by their birth and death values.

\subsubsection{TDA for neuroimage data}

TDA has emerged as a powerful framework for characterizing structural patterns in neuroimaging data, particularly in settings where classical intensity-based methods fail to capture complex morphology. As surveyed by \citep{singh2023tda}, TDA provides multiscale descriptors of shape and structure that are robust to continuous deformations and have been increasingly adopted in structural MRI analysis for both methodological development and clinical applications.

In structural MRI, one of the earliest and most widely used paradigms applies persistent homology directly to grayscale images, where voxel intensity values induce a sublevel- or superlevel-set filtration on a cubical complex \citep{franccois2024train,moon2023using, ali2025comparing, ahmed2026four} 
While intensity-based filtrations are conceptually simple and computationally convenient, they can be sensitive to imaging artifacts such as noise, scanner-dependent contrast variation, and intensity non-uniformity, all of which may distort filtration ordering and obscure biologically meaningful shape patterns~\citep{ganzetti2016intensity, turkevs2021noise}. These limitations have motivated topology-aware representations that emphasize anatomical structure rather than raw intensity values, particularly in settings where regions of interest can be delineated or when shape is the primary target of analysis~\citep{hu2019topology,singh2023tda}.

Building on this perspective, several structural MRI studies construct topology informed summaries directly from delineated anatomical or pathological regions and then integrate these summaries into downstream statistical models for prediction and inference. A representative example is the work of~\citep{crawford2020predicting}, who propose the \emph{smooth Euler characteristic transform} (SECT) as a functional topological descriptor of glioblastoma tumor morphology derived from MRI segmentations. Their approach is rooted in the Euler characteristic,
$\chi(A) = \sum_{k \ge 0} (-1)^k \beta_k(A),$
which aggregates Betti numbers into a single topological invariant. By evaluating the Euler characteristic across a family of directional sublevel-set filtrations and applying a smoothing operation, SECT yields a function-valued representation of tumor shape that is naturally compatible with functional regression. Using SECT-based predictors,~\citep{crawford2020predicting} demonstrate improved survival prediction relative to conventional geometric descriptors, illustrating how topology-aware functional summaries can support clinically relevant inference in structural neuroimaging. 
A prominent illustration of segmented structural MRI analysis using persistent homology is given by ~\citep{moon2023using}. Using segmented glioblastoma multiforme (GBM) subregions from contrast-enhanced T1-weighted MRI, they compute persistent homology on distance-based filtrations of the tumor masks to obtain multiscale summaries of morphological organization. The resulting persistence-derived descriptors capture clinically relevant patterns such as spatial fragmentation and ring-enhancing or necrotic structures that are difficult to quantify using conventional geometric or radiomic features. 

TDA can also be integrated directly into statistical modeling frameworks through stable functional embeddings of persistence diagrams. In particular,~\citep{moon2023using} propose representing persistence diagrams defined in Eq.~\ref{eq:PD}, using persistence surface functions,
\begin{equation}\nonumber
\rho_{\mathcal{P}}(x,y)
=
\sum_{(b,d)\in\mathcal{P}}
g_{(b,d)}(x,y)\, w(b,d),
\end{equation}
where $g_{(b,d)}(x,y)$ denotes a smoothing kernel centered at $(b,d)$ and $w(b,d)$ is a nonnegative weighting function that emphasizes persistent topological features. This functional embedding yields stable, noise-robust representations of multiscale topology and enables the use of standard tools from functional data analysis. Moreover, this study demonstrates how such topology-derived functional summaries can be incorporated into downstream inferential models alongside traditional clinical covariates. In survival analysis settings, they consider a functional Cox proportional hazards model of the form
\begin{equation}\nonumber
h_i(t)
=
h_0(t)\exp\left\{
\mathbf{Z}_i^\top \boldsymbol{\gamma}
+
\int X_i^{(0)}(u)\,\alpha(u)\,du
+
\int X_i^{(1)}(v)\,\beta(v)\,dv
\right\},
\end{equation}
where $X_i^{(0)}(\cdot)$ and $X_i^{(1)}(\cdot)$ denote functional summaries derived from 0- and 1-dimensional persistent homology, respectively, while $\alpha(\cdot)$ and $\beta(\cdot)$ quantify their associations with survival.

A complementary research direction leverages TDA more directly as a tool for anatomical segmentation itself. In this setting, persistent homology is not merely used to derive features, but rather to encode explicit morphological priors that guide segmentation of neuroanatomical structures. A recent example is the train-free segmentation framework proposed by~\citep{franccois2024train}, which introduces a fully unsupervised pipeline for MRI segmentation based on cubical persistent homology computed from superlevel-set filtrations. 
Persistent homology has also been used in neuroimaging to characterize structural brain variation associated with aging and neurodegenerative diseases. For example, \cite{saadat2021topological} demonstrated that \textit{Betti curves}, functional summaries of topological features computed across filtration thresholds, derived from structural MRI are predictive of age and can distinguish Alzheimer's disease (AD)-related structural alterations from normative aging trajectories. Complementing Betti curves, \emph{persistence landscapes}~\citep{bubenik2015statistical,bubenik2020persistence} provide stable functional representations of persistence diagrams, enabling their direct integration into statistical modeling frameworks through methods from functional data analysis~\citep{saadat2021topological,singh2023tda}.

While structural MRI provides a static snapshot of brain morphology, fMRI introduces an additional temporal dimension, capturing evolving patterns of neural activity over time. In fMRI analysis, topological methods are typically applied to time-resolved representations of brain activity, including correlation matrices, time-varying functional connectivity graphs, and point clouds constructed from multivariate  signals~\citep{ryu2023persistent,chung2024topological,bhattacharya2025persistent}. Persistent homology characterizes the evolving topological organization of the brain through summaries such as persistence diagrams, Betti curves, persistence landscapes, and other functional representations, providing powerful tools for studying dynamic brain states, cognitive state transitions, and inter-subject variability in functional organization~\citep{giusti2016two, saggar2018towards, das2023topological, gracia2020topological, martinez2023, catanzaro2024topological, santoro2024higher, wang2025topological}. 
Another TDA methodology used in fMRI research is the \textit{Mapper algorithm}~\citep{singh2007topological, madukpe2026comprehensive}.
Unlike persistent homology, which characterizes topological features across filtrations, Mapper produces a graph representation that encodes how clusters of similar observations are organized and connected in the data manifold. Applied to fMRI activation patterns, the resulting Mapper graph provides a topological skeleton of whole-brain dynamics, with nodes corresponding to recurrent activation configurations and edges reflecting their overlap and continuity in the underlying state space. 
To relate this topological summary to temporal organization, \cite{saggar2018towards} introduced a temporal connectivity matrix derived from the Mapper graph, which captures similarity between fMRI time frames rather than inter-regional coupling. Using this TDA based representation, they showed that whole-brain activation patterns exhibit structured mesoscale organization and that individual differences in the topology of these activation trajectories are predictive of task performance.

\subsubsection{TDA for brain networks}

Modern neuroimaging and neuroscience studies increasingly investigate brain organization at the network level, where complex interactions among distributed brain regions are represented using graph-based models~\citep{bullmore2009complex,rubinov2010complex}. Let \(G=(V,w)\) denote a brain network, where \(V\) is a set of nodes corresponding to brain regions obtained from a fixed parcellation and \(w_{ij}\) denotes a symmetric edge weight measuring structural or functional connectivity between regions \(i\) and \(j\). 
Classical graph-theoretic analyses typically rely on thresholded adjacency matrices and summarize network structure using single-scale metrics such as degree or clustering coefficient; however, these measures are highly sensitive to the choice of threshold~\citep{garrison2015stability,drakesmith2015overcoming}. Topological data analysis provides a principled multiscale alternative by considering a graph filtration
\[
G(\epsilon) = (V,E(\epsilon)), \qquad E(\epsilon)=\{(i,j)\in V\times V : w_{ij} \ge \epsilon\},
\]
which generates a nested sequence of graphs as the filtration parameter \(\epsilon\) varies. Persistent homology tracks changes in network topology across this filtration, and topological summaries, e.g., persistence diagrams and barcodes, provide a complete and stable multiscale characterization of brain network topology, forming the foundation for topology-aware comparison and learning of connectomic data \citep{songdechakraiwut2023topological, chung2024altered,chung2024topological}.

An illustrative example of this paradigm is provided by \citep{songdechakraiwut2023topological}, who demonstrate how persistent homology can be used to induce an explicit and interpretable alignment between brain networks. In their framework, network dissimilarity is quantified through \textit{Wasserstein distances}~\citep{wasserman2018topological,panaretos2019statistical} defined on the birth values of connected components and the death values of cycles, leading to the 0- and 1-dimensional (0D and 1D, respectively) topological losses
\[
L_{0\mathrm{D}}(G,P)
=
\min_{\tau_0}
\sum_{b \in I_0(G)}
\bigl( b - \tau_0(b) \bigr)^2,
\qquad
L_{1\mathrm{D}}(G,P)
=
\min_{\tau_1}
\sum_{d \in I_1(G)}
\bigl( d - \tau_1(d) \bigr)^2,
\]
and their combined topological loss
\[
L_{\mathrm{top}}(G,P) = L_{0\mathrm{D}}(G,P) + L_{1\mathrm{D}}(G,P).
\]

In addition, \cite{chung2024altered} employ persistent homology to study alterations in white matter structural covariance networks in maltreated children. Building on graph filtrations and Betti curve representations, the authors construct weighted structural covariance networks from tensor-based morphometry and diffusion MRI, and summarize their multiscale topology using Betti curves. To enable statistical inference, the authors introduce a global test statistic based on the separation between within-group and between-group topological distances. Specifically, given two groups of networks, the total within-group distance is defined as
\[
\ell_W = \sum_{i,j} d(X_i, X_j) + \sum_{i,j} d(Y_i, Y_j),
\]
while the between-group distance is defined as
$\ell_B = \sum_{i,j} d(X_i, Y_j),$
where \(d(\cdot,\cdot)\) denotes the Wasserstein distance between networks computed from their persistent homology summaries ($X$ and $Y$).

Group differences in network topology are then assessed using the ratio statistic
$\phi = {\ell_B}/{\ell_W},$
which captures the extent to which topological variability between groups exceeds variability within groups. This formulation leverages the multiscale nature of persistent homology to provide a single global test over all filtration values, avoiding multiple-comparison issues inherent in threshold-based graph analyses. Applied to white matter structural covariance networks, this approach revealed that maltreated children exhibit fewer connected components and more densely correlated network structure across scales, highlighting how TDA can uncover clinically meaningful alterations in brain network topology that are not readily detected by conventional graph-theoretic metrics.

A natural next step is to use these distances to organize heterogeneous clinical cohorts without relying on unstable single-threshold summaries. \cite{xu2024topology} propose a topology-based clustering framework for resting-state functional brain networks in an AD cohort, where each subject's connectome is first mapped to persistent homology features across a graph filtration and then compared using a topological distance. Concretely, given two subjects \(i\) and \(j\), a pairwise similarity kernel is constructed from the topological distance \(d_{\mathrm{top}}(i,j)\) (computed from persistence-based summaries), e.g.,
$K_{ij}=\exp\{-\gamma\, d_{\mathrm{top}}(i,j)\},$ 
so that subjects with similar multiscale connectivity topology have larger affinity. Clustering is then performed in this topology-induced similarity space (e.g., via spectral clustering), yielding data-driven subtypes whose separation is driven by persistent topological structure rather than by arbitrary graph thresholds or a small set of handcrafted network statistics. 
Extending these ideas further, \cite{das2023topological} introduce an order-statistics framework to simplify and statistically formalize persistent homology computations on human brain networks, allowing for population-level inference while accounting for randomness and heterogeneity in connectomes. In contrast to computing barcodes directly from each weighted network, the edge weights \(W_1,\ldots,W_q\) of a complete graph with \(q=\frac{p(p-1)}{2}\) edges can be treated as random variables and ordered as
\[
W_{(1)} \le W_{(2)} \le \cdots \le W_{(q)},
\]
where \(W_{(k)}\) denotes the \(k\)th order statistic of the edge weights. The authors leverage these ordered statistics to derive expected birth and death values for connected components and cycles under a specified distribution of edge weights, yielding  \textit{expected persistent barcodes}~\citep{Carlsson_2004b} that characterize the typical topology of networks across a population. By comparing these expected topological summaries between groups, they demonstrate statistically significant differences between male and female resting-state functional brain networks.

\subsection{Knowledge Graph representations for Clinical Interoperability}
\label{Sec:KJ:KG}
In a medical response environment, first responders need to be able to exchange information, queries, and requests with some assurance that they share a common meaning. This interoperability requirement is not just for the data itself but also for describing their policies for sharing data and valid usage. One possible approach to this issue is to employ \textit{Semantic Web} techniques for modeling and reasoning about the data and the information related to image analytical models. A standard way of employing semantic web techniques is to use knowledge graphs or ontologies and reasoning network models according to the rules in policy texts. The semantic web enables data to be annotated with machine-understandable meta-data, allowing the automation of its retrieval and usage in the correct context. Semantic web technologies include languages such as Resource Description Framework (RDF) \cite{pan2009resource} and Web Ontology Language (OWL) \cite{antoniou2009owl} for defining ontologies and describing meta-data using these ontologies as well as tools for reasoning over these descriptions. 

Researchers have modeled access control policy reasoning using various semantic web languages \cite{kagal2002rei, uszok2008new, gicquel2023survey, antoniou2004semantic, gupta2023ontology, uszok2008new, chattoraj2024semantically}.  These technologies can be used to provide common semantics of service information and policies enabling software agents who understand underlying technologies to communicate and use each other's data and services effectively.  OWL has a well-defined semantics grounded in first order logic and model theory, allowing programs to draw inferences with the assurance that the subsequent interpretation is sound. An important advantage for OWL over many other knowledge-representation systems is that it has well-defined subset profiles guaranteeing sound and complete reasoning with various levels of reasoning complexity and designed to work with popular implementation technologies, such as OWL QL for databases and OWL RL for rule-based systems.

\subsubsection{Knowledge Graphs in Translational Neuroscience}
Translational neuroscience aims to convert mechanistic discoveries at the molecular and genetic scale into clinically actionable diagnostics and therapies, a process that hinges on integrating evidence otherwise fragmented across publications, ontologies, and specialised repositories. Knowledge Graphs (KGs) address this fragmentation by representing biomedical entities, proteins, molecular pathways, anatomical structures, drugs, phenotypes, and diseases as nodes linked by semantically labelled edges, yielding a unified and queryable substrate for reasoning across scales \cite{Chattoraj2025LLMBK, Chattoraj2024MedRegKGKF}. Large-scale biomedical KGs demonstrate the translational payoff of this representation integrating knowledge from millions of studies to prioritise drug response candidates via link prediction over compound gene disease paths \cite{himmelstein2017systematic}, while PrimeKG assembles a multimodal precision-medicine graph supporting disease-centric machine learning via spanning ten biological scales from disease-associated protein perturbations to phenotypes and approved therapies. \cite{chandak2023building}.

In neuroscience specifically, neurodegenerative and psychiatric disorders are inherently multi-scale. Alzheimer's disease (AD), for
example, couples genetic risk (e.g., \textit{APOE}), amyloid and tau
proteinopathy, regional atrophy and network disruption, and progressive
cognitive decline. Encoding these entities and their relations in a graph enables patient stratification, mechanistic hypothesis generation, and biomarker disease association discovery. These entities and relationship provides interpretable relational paths which is an increasingly important property for clinical and
regulatory acceptance \cite{chattoraj2024semantically, yang2023kgxdp}. The Conventional knowledge graphs represent information using dyadic (pairwise) relations, where each edge connects exactly two entities. However, many neurobiological processes involve inherently higher-order interactions among multiple biomarkers, genes, proteins, and brain regions that cannot be fully captured by pairwise relationships alone \cite{battiston2020networks,benson2018simplicial}. The higher-order biological relationships reduction to pairwise edges obscures the joint interactions among multiple entities, preventing their explicit representation and potentially limiting the ability to capture biologically meaningful diagnostic and mechanistic patterns. This limitation motivates the use of hypergraphs, whose hyperedges directly model relationships among arbitrary numbers of entities \cite{zhou2006learning}.

\subsubsection{Hypergraph-Structured Representations for Neuroimaging Cohorts}
The hypergraph generalises a graph by allowing each edge to connect an arbitrary number of nodes. Formally, a weighted hypergraph is $\mathcal{G}=(\mathcal{V},\mathcal{E},\mathbf{W})$, where $\mathcal{V}$ is a set of vertices, $\mathcal{E}$ is a set of hyperedges with each $e\in\mathcal{E}$ a non-empty subset $e\subseteq\mathcal{V}$, and $\mathbf{W}$ assigns each hyperedge a weight. The structure is captured by an incidence matrix $\mathbf{H}\in\{0,1\}^{|\mathcal{V}|\times|\mathcal{E}|}$ with
$\mathbf{H}(v,e)=1$ iff $v\in e$; an ordinary graph is the special case in which $|e|=2$ for all $e$. 

Spectral learning on hypergraphs, and hyperedge convolution in hypergraph neural networks (HGNN), exploit this structure to propagate information along higher-order relations and to fuse multiple data modalities within a single model \cite{zhou2006learning} \cite{feng2019hypergraph}. In neuroimaging cohorts, this capability has been used to diagnose AD from incomplete multi-modality data by
aligning modality-specific hypergraph views \cite{liu2017view}, illustrating a natural fit between hypergraph learning and the multimodal, partially-missing data that characterise studies such as the Alzheimer's Disease Neuroimaging Initiative (ADNI) \cite{jack2008alzheimer}.

\begin{table}[t]
\centering
\caption{Vertex (node) types in the ADNI hypergraph and their instantiation
in ADNI variables.}
\label{adni-hg-nodes}
\small
\begin{tabularx}{\linewidth}{@{}l >{\raggedright\arraybackslash}X@{}}
\midrule
\textbf{Node type} & \textbf{ADNI instantiation} \\
\midrule
Subject & Participant, labelled CN / MCI / AD \\
Brain ROI & FreeSurfer-parcellated regions from T1 MRI (hippocampus,
                       entorhinal cortex, posterior cingulate, precuneus) \\
Imaging measure & Structural MRI volume/thickness, FDG-PET, amyloid-PET
                       (\textsuperscript{18}F-florbetapir), tau-PET
                       (\textsuperscript{18}F-flortaucipir) \\
Fluid biomarker  & CSF A$\beta$42, t-tau, p-tau181; plasma p-tau \\
Genetic marker & APOE~$\varepsilon$4 allele count; candidate SNPs;
                       polygenic risk score \\
Cognitive / clinical & MMSE, ADAS-Cog, CDR-SB, RAVLT \\
Demographic & Age, sex, education \\
Visit & bl, m06, m12, m24 \\
\midrule
\end{tabularx}
\end{table}

\begin{table}[t]
\centering
\caption{Hyperedge (higher-order relation) types in the ADNI hypergraph.}
\label{adni-hg-edges}
\small
\begin{tabularx}{\linewidth}{@{}>{\raggedright\arraybackslash}p{0.24\linewidth}
                                >{\raggedright\arraybackslash}X
                                >{\raggedright\arraybackslash}X@{}}
\midrule
\textbf{Hyperedge} & \textbf{Members it connects} &
\textbf{Higher-order rationale} \\

$e_1$ Multimodal biomarker profile &
Subject $+$ its discretised imaging, fluid, genetic, and cognitive markers at a visit &
Diagnosis is the joint configuration, not any single pair \\[2pt]

$e_2$ Subject similarity (kNN) &
A subject and its $k$ nearest neighbours in one modality's feature space &
Multimodal hypergraph $=$ union over modality-specific hyperedges \\[2pt]

$e_3$ Regional co-atrophy / co-activation &
A set of ROIs that jointly deviate (e.g., default-mode network) &
Captures network-level dysfunction across many regions at once \\[2pt]

$e_4$ Longitudinal trajectory &
The same subject's states across visits &
Encodes progression CN~$\rightarrow$~MCI~$\rightarrow$~AD \\[2pt]

$e_5$ Mechanistic bridge &
Genetic, molecular, anatomical, and disease nodes (e.g., APOE--amyloid
pathway--medial-temporal ROIs--AD) &
Links the cohort hypergraph to an external biomedical KG \\
\midrule
\end{tabularx}
\end{table}

\begin{figure*}[!t]
\centering
\includegraphics[width=0.5\textwidth]{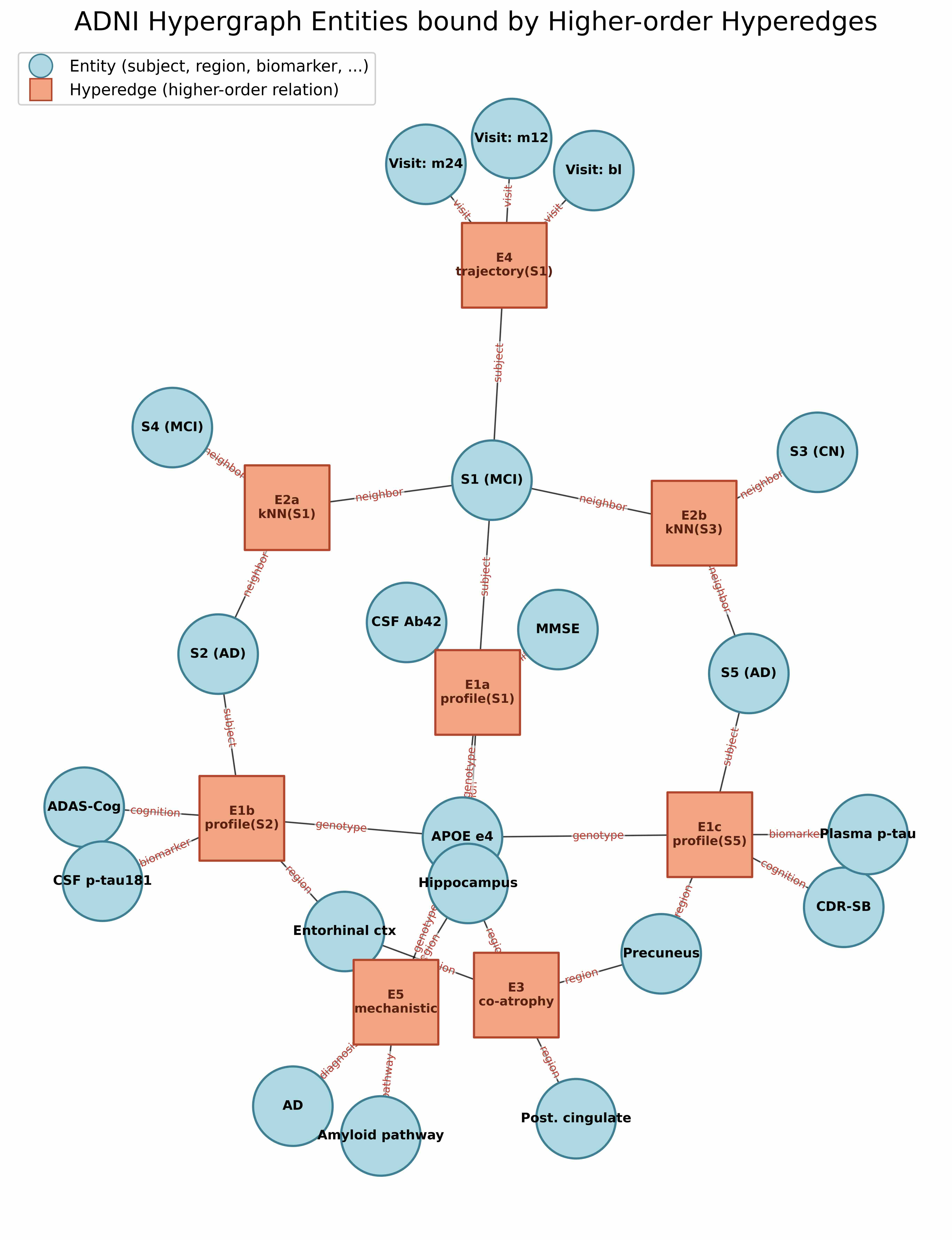}
\caption{Schematic illustration of a hypergraph representation over ADNI entity types; nodes and memberships}
\label{hypergraph}
\vspace{-3mm}
\end{figure*}

Let each ADNI participant, biomarker, region, and assessment be a vertex, and let higher-order clinical and biological relations be hyperedges. Vertex types $\mathcal{V}$ and hyperedge types $\mathcal{E}$ are summarised in Table~\ref{adni-hg-edges}, \ref{adni-hg-nodes}. A canonical \textit{multimodal biomarker profile}
hyperedge for subject $s$ at visit $t$ takes the form:
\begin{equation}
  e_{\text{profile}}^{(s,t)}=\bigl\{\,s,\ \mathrm{Hipp}^{\downarrow},\
\mathrm{A}\beta 42^{\downarrow},\ \mathrm{APOE}\text{-}\varepsilon 4^{+},\
\mathrm{MMSE}^{\downarrow}\,\bigr\}  
\end{equation}
encoding the AD signature as the co-occurrence of medial-temporal atrophy, amyloid positivity, genetic risk, and cognitive impairment, information that no set of pairwise edges preserves without loss. In Fig. \ref{hypergraph} the schematic illustration of a hypergraph representation over ADNI entity types; nodes and memberships is illustrated.

%%%%%%%%%%%%%%%%%%%%%%%%%%%%%%%%%%%%%%%%%%%%%%%%%%%%%%%%%%%%%%%%%%%%%%%%%%%%%%

\section{Artificial intelligence techniques for neuroscience studies}
\label{Sec:AI_Neuro}

\subsection{AI for neuroimage modeling}

Artificial intelligence has become central to the analysis of neuroimage data in neurodegenerative disease, where early, individualized detection is critical but structural neuroimaging and clinical assessment alone often identify disease only after substantial neuronal loss \citep{kwon2025coordinated, zhou2025generative}. Deep learning models can learn hierarchical features directly from high-dimensional, multimodal clinical data --- imaging, speech, gait --- without hand-engineered feature extraction \citep{chaki2023deep, olaniyan2023new, sharma2023deep, bossa2024generative}, and generative AI increasingly complements this with synthetic data generation, augmentation, and disease modeling. This review assumes familiarity with the underlying machine-learning and deep-learning concepts (CNNs, RNNs, GANs, diffusion models, transformers, and so on) and instead concentrates on how these tools are actually being used in neuroimage analysis and neuroscience research: applied diagnosis and progression modeling for Alzheimer's disease (AD) and Parkinson's disease (PD), recent large-scale foundation models, generative AI for synthesis and modeling, the emerging paradigm of AI-enabled personalized digital brain twins, and the limitations and open problems that cut across this literature.

\subsection{Deep Learning for Diagnosis and Progression Modeling of Neurodegenerative Disease}
Volumetric convolutional architectures remain the workhorse of neuroimaging-based AD/PD diagnosis.  \cite{dhinagar20213dcnn} distinguished PD and AD directly from T1-weighted MRI volumes with a 3-D CNN, and  \cite{islam2018brain} used an ensemble of CNNs for multi-stage AD detection with reliable performance across disease severity. For PD specifically, transfer learning from pretrained architectures (VGG, ResNet, DenseNet) has proven useful given typically small disease-specific cohorts \citep{khan2024detection};  \cite{alissa2021parkinsons} built an early-diagnosis system around a figure-copying motor task, and  \cite{mozhdehfarahbakhsh2021mri} forecast PD progression stages directly from MRI. This CNN-centric literature has grown very large, and a substantial parallel body of systematic and narrative reviews now synthesizes it \citep{gao2023review, toumaj2024applications, srinivasan2025future}; a recurring observation across this survey literature is that reported accuracies, while often high, are typically obtained on a single, comparatively small, single-site cohort, foreshadowing the reproducibility and generalization concerns.

Recurrent and hybrid architectures address the temporal and multimodal character of disease progression rather than a single static scan  \cite{balaji2021automatic, miltiadous2023dice, lee2019predicting}. Deep networks that fuse clinical, genetic, and imaging modalities \citep{segovia2018using, basher2021volumetric} tend to outperform single-modality, single-architecture models: hybrid CNN--LSTM systems have reported higher AD-detection accuracies \citep{zolfaghari2025hybrid}, multimodal CNN--LSTM frameworks integrating MRI and PET jointly capture structural and temporal disease signatures \citep{shaikh2025multimodal}, and RNN-based architectures have been used for simultaneous AD and PD detection from brain MRI \citep{saranraj2025sequential}. A representative recent example, NeuroNet, couples a 3-D CNN, an LSTM, and an attention-based region-of-interest selection mechanism with Bayesian hyperparameter optimization to reach comparable accuracy on ADNI data \citep{golovanevsky2022multimodal}.

A closely related but distinct problem is detecting and quantifying \emph{change} between longitudinal scans of the same individual, which predates deep learning and has an extensive classical image-processing lineage --- registration-based subtraction, statistical change-vector analysis, and dictionary-learning or clustering-based techniques --- subsequently extended to specific neurological and demyelinating applications such as diffusion MRI and multiple sclerosis lesion tracking. Deep-learning approaches now recast longitudinal change detection as representation learning: joint autoencoders and Siamese networks model unsupervised change and continuous disease-severity trajectories from paired images \citep{dufresne2020joint, li2020siamese}, deep metric learning directly compares longitudinal scans \citep{pmlr-v227-kim24a}, and longitudinal pooling with consistency regularization models progression directly from MRI sequences \citep{ouyang2020longitudinal}. Neural ODEs have been used to learn an explicit spatio-temporal model of disease progression from longitudinal volumetric data \citep{lachinov2023learning}, temporally aware diffusion models connect this literature to generative modeling \citep{litrico2024tadm}, general-purpose multimodal large language models have been evaluated for identifying radiologic progression directly from serial brain MRI \citep{kelly2025can}. Deep learning has further been applied to large-scale histological tau protein mapping, bridging macroscopic imaging-based progression tracking with microscopic validation of the underlying pathology \citep{alegro2019deep}.

\subsection{Foundation Models and Generative AI for Neuroimaging}
A recent and fast-moving development is the emergence of large-scale \emph{foundation models} pretrained via self-supervision on broad, aggregated neuroimaging and neurophysiological corpora, intended as general-purpose backbones adaptable across downstream tasks via fine-tuning or zero-shot inference. BrainLM, pretrained on 6{,}700 hours of aggregated fMRI, supports both fine-tuned clinical-variable prediction and zero-shot identification of intrinsic functional networks from learned attention patterns \citep{caro2024brainlm}. Brain-JEPA instead pretrains via a joint-embedding predictive objective with a specialized spatial encoding of cortical/subcortical topology \citep{dong2024brainjepa}, and SwiFT applies a hierarchical shifted-window transformer directly to raw 4-D fMRI, serving both as a supervised phenotype-prediction architecture and as a backbone for subsequent pretraining \citep{kim2023swift}. A parallel line of foundation models targets EEG and intracranial recordings, motivated by their larger aggregate public data volume and clinical ubiquity in epilepsy, sleep, and brain--computer interfaces: LaBraM combines a vector-quantized neural tokenizer with masked-EEG-modeling pretraining across roughly 2{,}500 hours of data \citep{jiang2024labram}, NeuroLM bridges EEG and natural-language representations within a unified multi-task model \citep{jiang2024neurolm}, BrainBERT applies BERT-style masked pretraining to intracranial recordings \citep{wang2023brainbert}, and EEGPT targets universal, cross-dataset EEG representation learning \citep{wang2024eegpt}. 

Generative AI --- chiefly GANs and, more recently, diffusion models --- has attracted considerable applied interest for synthetic neuroimage generation, augmentation, and disease-progression modeling in AD and PD \citep{wang2023applications, bacon2024neuroimage, cheong2025survey, hatem2025ai, yang2025genbrain, beste2026generative, parmar2026towards}. GAN and VAE architectures have been used to generate synthetic AD MRI addressing class imbalance and rare-subtype limitations, AD progression via adaptive-regression, ROI-enhanced design,  learning latent representations predictive of amyloid-related AD progression dynamics  and large-scale architectures \citep{qiao2020biggan, bharathi2024aregan, bossa2024generative}. Beyond diagnosis, neuroplasticity-inspired GANs have been used to simulate functional brain regeneration and adaptive cognitive recovery \citep{khajgiwale2025neuroplasticity}, and federated GAN training has enabled multi-institutional AD classification from EEG under data-privacy constraints \citep{stephen2026federated}, while large language models and multimodal chatbot systems are beginning to be explored for AD disease-literacy and communication applications \citep{bosco2025testing}. Diffusion models have followed closely, valued for training stability and sample quality relative to GANs: large-scale latent diffusion has generated high-resolution synthetic structural brain MRI \citep{pinaya2022brain}, semi-supervised diffusion has been used for brain-age prediction \citep{ijishakin2024semi}, diffusion-generated synthetic data has improved downstream prediction tasks \citep{hu2024synthetic}, and normative representation learning within generative models has been evaluated for anomaly detection \citep{bercea2025evaluating}. Connectome- and multimodality-aware diffusion architectures such as ConnectomeDiffuser \citep{chen2025connectome} and successive versions of BrainNetDiff \citep{zong2024brainnetdiff, zong2024brainnetdiff}, together with adaptive, clinical-aware latent diffusion for missing-modality imputation \citep{zhou2026adaptive}, extend generative modeling toward the multimodal-fusion problems.

\subsection{AI, Digital Twins, and Digital Clones of the Brain}
A distinct and rapidly growing application of AI in neuroimaging is the construction of \emph{digital twins} (sometimes called digital clones) of the brain: personalized, generative, computational models calibrated to an individual's own imaging data and intended to simulate, predict, and eventually help steer that individual's brain function.  \cite{wang2024virtualbraintwins} formalize this ``virtual brain twin'' concept, describing a pipeline in which subject-specific structural and functional imaging data are used to assemble individualized brain geometry and connectivity, calibrate a personalized whole-brain model, and support clinical inference across healthy ageing and several diseases, including Alzheimer's and Parkinson's disease; this builds on earlier infrastructure such as The Virtual Brain platform for connectome-constrained whole-brain simulation \citep{ritter2013}.  \cite{zimmermann2022} similarly frame brain digital twins as a broader computational-revolution concept for clinical neuroscience, emphasizing their potential for individualized diagnosis, prognosis, and treatment planning rather than population-averaged inference.

AI enters this pipeline at essentially every stage. Deep-learning-based segmentation, registration, and parcellation supply the personalized anatomical and connectivity inputs on which a digital twin is built. Foundation models  offer pretrained, transferable representations that can be fine-tuned to an individual's own scans with comparatively little subject-specific data, directly addressing the small-sample calibration problem that would otherwise limit personalization; and generative models --- GANs, VAEs, and diffusion models --- are increasingly used both to synthesize plausible individualized brain data for calibration and validation and, more ambitiously, as learned surrogates that approximate the output of a computationally expensive mechanistic whole-brain simulation at a fraction of the cost, enabling large-scale virtual-cohort studies and near-real-time clinical use. A concrete illustration of AI-enabled digital-twin methodology in a clinical setting is provided by  \cite{zhang2026}, who use a personalized digital-twin-brain framework to identify state-specific neurostimulation targets for abnormal brain dynamics in tinnitus, directly linking individualized computational modeling to a candidate therapeutic intervention.

Despite this promise, AI-driven digital twins of the brain inherit, and in some respects amplify, the limitations discussed below: personalization to a single subject's sparse imaging data raises acute identifiability and overfitting concerns, deep-learning surrogates trained to mimic mechanistic simulations require careful validation against the mechanistic model they replace rather than against clinical outcomes alone, and the clinical and regulatory validation required before an individualized, AI-calibrated brain model can inform real treatment decisions remains largely unresolved. Digital twins therefore represent both one of the most clinically compelling and one of the most methodologically demanding current applications of AI to neuroimage data.

\subsection{Change detection in medical images}
\label{Sec:VS:ChangeDetection}

Medical imaging is essential for diagnosing disease, planning treatment, and continuous disease monitoring. However, as medical images become increasingly abundant in volume and complicated in structure, there arises the need to automate processes that analyze medical images. In this context, the task of change detection aims to detect differences between successive images, reducing the burden of manual comparison, increasing inter-observer agreement, and focusing attention on clinically meaningful differences \citep{patriarche2007change, lindquist2007changepoint}. The central issue in medical image change detection is the challenge of separating genuine pathologies from artifacts caused by factors such as changes in patient positioning, variations in imaging protocols, and natural physiological deformation of organs \citep{patriarche2004review}. Early solutions involved elaborate  techniques of image registration and normalization \citep{venot1984digital} followed by digital subtraction to compare registered images and find changes through analyzing difference images. Current change detection technologies serve various clinical goals, including tracking the course of diseases, assessing therapy response, monitoring recurrence, and quantitatively measuring disease evolution for clinical trials\citep{patriarche2007change}. Over the years, the area has progressed from basic image differences to machine learning-based approaches that are able to predict disease trajectory and anticipate future anatomical alterations\citep{lachinov2023learning, ouyang2020longitudinal}.

\subsubsection{Classical techniques for image change detection}

Fundamental to change detection is image registration to align images, eliminating positional differences. In \citep{ettinger1994automatic},  an energy minimization algorithm was proposed with three phases of refinement for image registration. Difference images produced by digital subtraction reveal changes in the images analyzed\citep{venot1984digital}. However, this method suffers from sensitivity to noise and inaccuracies. A more sophisticated strategy models the distribution of intensities in difference images. Statistical change detection approaches offer rigorous theoretical bases due to their consideration of inherent variability in medical images. \citep{halder2019change} proposed change vector analysis and  clustering techniques to handle ambiguity around boundaries. In cases where training sets are insufficient, unsupervised learning algorithms can prove effective. \citep{mazzei2023change} Proposed an approach combining feature extraction, multivariate analysis, and PCA. 
Besides changes between successive images, anatomical structures may also deform during imaging. There exist automatic approaches for detecting anatomical structures, measuring deformations, and distinguishing physiological growth from disease-caused changes\citep{zhou2006systems}. This allows separation of growth from pathological change. In this context, \citep{naitsat2017differential} proposed an approach using differential geometry defined on tetrahedral meshes to quantify deformation necessary to deform images to a canonical representation, illustrated by applications on tumors and anatomical feature clustering.

\subsubsection{Deep learning for change detection}
\label{Sec:VS:ChangeDetection_DL}

Recent advancements in deep learning have made possible change detection pipelines in which image features are learned entirely automatically. For example, \citep{nika2014change} proposed an approach that fuses locally-trained dictionary representations and PCA in order to detect changes in brain MRI images while minimizing the effects of positional variation.  An alternative deep learning architecture for comparing images is the siamese neural network for learning similarities; \citep{li2020siamese} developed a convolutional siamese neural network that predicts disease severity and classifies changes, exhibiting excellent generalization abilities for detecting disease progression and regression for retinopathy and osteoarthritis patients. In \citep{pmlr-v227-kim24a} an architecture (PaIRNet) was proposed for  longitudinal image pair  inputs for detecting significant changes using shared feature extractors and ranking layers. Here, the network trains by both predicting changes and ordering image sequences temporally via self-supervised learning. Illustrative examples are  in brain cancer, AD, and embryo development applications.

The inherent sequential nature of longitudinal medical imaging gives rise to the temporal sequence modeling methods.  In \citep{ouyang2020longitudinal},  a new type of pooling layer called longitudinal pooling  was used to predict progressive changes in AD, alcohol use disorder, and adolescent neurodevelopment. 
When there is a lack of longitudinal medical image labels, \citep{to2021selfsupervised} developed an unsupervised approach for generating synthetic alterations in lesions for generating pseudo-labels; this technique performs well for lesion detection and localization in patients suffering from multiple sclerosis.  Similarly, \citep{emre2022tinc} adapted non-contractive learning via a temporal similarity loss, allowing the prediction of disease conversions from retinal optical coherence tomography (OCT) volume images.

More recently, continuous-time models have been adopted for predicting long-term disease progression. For example, \citep{lachinov2023learning} proposed a continuous change prediction scheme where baseline images are converted into initialization parameters for an ordinary differential equation (ODE), where  the solution predicts continuously changing states and segmentation for geographic atrophy and AD.

\subsection{Limitations}
Neuroimaging AI faces limitations that are unusually severe relative to data-rich AI domains. Sample sizes remain small relative to feature dimensionality: cross-validated accuracy estimates at typical neuroimaging sample sizes carry substantial uncertainty \citep{varoquaux2017}, and reproducible brain--behavior associations have been shown to require far larger cohorts than the median published study uses \citep{marek2022}; consistent with this, deep networks offer only modest gains over well-tuned classical models at realistic sample sizes \citep{li2019deep, he2020, schulz2020}, motivating the large-scale pretraining strategies. Data leakage --- improper cross-validation splitting or feature selection performed outside the validation fold --- has repeatedly been documented to inflate reported neuroimaging classification accuracies \citep{poldrack2019, poldrack2020, kapoor2023}, echoing longer-standing statistical reproducibility failures such as inflated cluster-inference false-positive rates \citep{eklund2016}. Interpretability remains a persistent concern: post hoc explanation methods such as Grad-CAM and SHAP \citep{selvaraju2017, lundberg2017} applied to the \emph{same} trained model can produce substantially divergent brain maps \citep{thomas2022interpret}, a concern amplified for the zero-shot, emergent behaviors of large foundation models. Models trained at one site or on one scanner frequently generalize poorly elsewhere \citep{arbabshirani2017}, aggregated pretraining corpora remain unevenly distributed across scanners and populations, and foundation-model pretraining carries substantial computational cost \citep{bommasani2021}. Finally, strong cross-validated predictive accuracy does not by itself establish biological or causal validity, or robustness to the diagnostic-criterion shifts common in clinical practice \citep{jack2018, bzdok2018}.

\subsection{Open Problems and Future Directions}
Several open problems follow directly from these limitations. Neuroimaging-specific scaling laws for foundation models such as BrainLM, Brain-JEPA, and LaBraM remain poorly characterized, and it is unclear what pretraining scale would close the deep-learning-versus-classical-method performance gap noted above \citep{brown2020gpt3}. Interpretability methods need to move from merely \emph{available} to demonstrably \emph{faithful}, with architecture-level interpretability constraints such as BrainGNN's region-selection pooling \citep{li2021braingnn} offering one promising direction. Multimodal fusion --- combining volumetric, connectome, and genetic/phenotypic data within a single foundation-model framework, building on early multimodal autoencoder \citep{suk2014} and cross-modality GAN imputation \citep{pan2018} efforts --- remains underdeveloped relative to unimodal pretraining. Federated and privacy-preserving learning offers a route to multi-site pretraining scale without centralized data pooling, directly relevant to both foundation-model training and digital-twin personalization. More fundamentally, the field needs to move from purely correlational prediction toward causal and mechanistic understanding \citep{bzdok2018}, including tighter integration of learned representations with explicit biophysical and dynamical-systems models, and continued development of unsupervised subtyping methods for heterogeneous clinical populations \citep{kernbach2022}. Finally, translating this progress into deployed clinical tools --- including AI-calibrated digital twins --- will require biologically grounded validation frameworks \citep{jack2018, atalay2023digital, badano2023stochastic} and regulatory standards comparable to those emerging for AI in biomedicine more broadly \citep{esteva2019}, while preserving the individual-level predictive precision that motivated this literature in the first place \citep{finn2015}.

\section{Dynamical systems and mechanistic modeling of neuroimage data}
\label{Sec:DynSys_Neuro}

The brain is a dynamical system: neural activity, functional connectivity, and --- in the context of neurodegenerative disease --- the spatial distribution of pathological protein aggregates and tissue atrophy all evolve continuously over time, and their temporal evolution is shaped by the underlying anatomical connectivity of the brain. Much of quantitative neuroscience has instead relied on statistical approaches that are inherently \emph{inferential} and largely focused on the analysis of stationary patterns of localized activity: such analyses can reveal which brain region or network mediates a given cognitive process, but not the time evolution of that mediation, nor the causal or controlling mechanism responsible for it. It was argued in \cite{john2022} that this static, local approach makes it difficult to link neuroimaging results to plausible underlying neural mechanisms, and that dynamical systems theory provides the mechanistic framework needed to characterize both the brain's time-varying quality and its partial stability under perturbation.

Dynamical systems of neuroimaging data can be typically cast as nonlinear, stochastic state-space models. At the finest level of description, single-neuron (biophysical) models aim to replicate the electrophysiological properties of individual neurons \citep{hodgkinhuxley1952, fitzhugh1961, nagumo1962}.
Such models have been applied in understanding neurodegeneration, for example, on how amyloid-$\beta$-mediated disruption of glutamatergic synaptic transmission alters the firing rhythm of hippocampal dentate-gyrus--CA3 circuits \citep{dong2022glutamatergic} and various AD studies \citep{jiang2020dynamics, perez2016analyzing, mittag2023modelling}. Extensions to spatially continuous frameworks appear in \cite{you2023memristive}.

Neural mass models describe the average activity of large populations of neurons using a small number of coupled differential equations per cortical region \citep{wilson1972, jansen1995, david2003, moran2007}. Their usage in neuroscience and neurodegenerative disease studies appear in 
\cite{jirsa1996, breakspear2003, deco2008, cabrera2025fluctuations}. 
Whole-brain network models couple a neural mass model at each of several anatomically defined brain regions through the empirically measured structural connectome, typically obtained from diffusion MRI tractography \citep{honey2007, honey2009, ghosh2008, deco2011, cabral2011, deco2013, ritter2013, sanzleon2015, breakspear2017, cabral2017}. Prominent examples of such models include the Hopf (Stuart--Landau) model  \cite{deco2017hopf} and the R\"ossler  model \citep{rossler1976, piccinini2021}.  These models capture nonlinear effects   such as multistability, excitability, and oscillations across scales—from ion channels and spikes to  whole-brain networks. The models are mechanistic, though largely phenomenological, where states and parameters correspond to physiological quantities such as coupling strengths, time constants, and, when relevant, activity-dependent inputs. These mechanistic forward models capture causality and have the capacity to inform design of experiments, hypothesis generation and testing, and causal inference. Multiple studies have applied whole-brain models to neurodegenerative diseases \citep{patow2022whole,  stefanovski2019linking}. Some recent studies have also explored human behavior through models of the whole-brain dynamical systems. This includes relating dynamical systems concepts to neuroimaging data of meditation and sleep   \cite{deco2019, galadi2021}, as well as various clinical and psychiatric conditions such as epilepsy, migraine,  autism and depression \cite{dahlen2013, jirsa2019, iravani2021, wang2022,  he2025, wu2025, zhang2026}.

Additional studies leveraging dynamical systems for understanding neurodegenerative diseases appear in \citep{kuhn2008high, ray2008local, stoffers2008dopaminergic, stam2009graph,  bhattacharya2011alpha, rubin2012basal, hsiao2013altered, pavlides2015computational, koelewijn2017alzheimer, perez2016analyzing, bruna2023meg,  mittag2023modelling}. In a different direction, \textit{dynamic causal modeling (DCM)} \citep{friston2002dcmestimation, friston2003dcm,  daunizeau2009, stephan2009, li2011}, formulated a specific, low-dimensional instance of the state-space framework  for hypothesis-driven analysis of task-based effective connectivity. 

 Unlike the neural-activity models that typically operate on a timescale of milliseconds to seconds, disease-progression models operate on a timescale of months to decades. However, neurodegenerative diseases, despite their diverse clinical presentations, share the hallmark accumulation of misfolded protein aggregates that emerge in specific, disease-characteristic brain regions and subsequently progress along anatomical (synaptic and axonal) networks \citep{alexandersen2026network, frost2010prion, jucker2013self, braak1991neuropathological, liu2012trans, vogel2020spread, raj2012, raj2015, zhou2012, jucker2013self, prusiner1998, frost2010prion} and numerous others. 
 A  complementary class of models addresses a version of the disease-progression problem in which the explicit differential-equation dynamics  are replaced by a statistical model of the \emph{ordering} or \emph{staging} of biomarker abnormality, particularly useful when only cross-sectional (rather than densely longitudinal) data are available and the true chronological disease onset time for each individual is unknown \citep{fonteijn2012, jedynak2012, young2018}.  However, as a caveat,   note that \cite{rollo2023dynamical} observed that AD is fundamentally a spatio-temporal dynamic pathology unfolding across genetic, cellular, tissue, and organ-level scales simultaneously.

% #####################################################################
\subsection{Statistical and machine learning tools for learning dynamical systems}
% #####################################################################

For state-space models that underlie many dynamical systems used in neuroscience,  the Kalman filter and its non-trivial modern extensions \citep{kalman1960, ghahramani1999} provides recursive Bayesian approximation for the posterior distribution of the latent state  given data. For learning disease-progression dynamical systems (as opposed to neural-activity dynamical systems), a framework has been developed by \citep{garbarino2021investigating}. Other relevant studies are by \cite{ryali2011, iturriamedina2016, durstewitz2017, chen2018node, iturria2018multimodal, pandarinath2018,   koppe2019, stefanovski2019linking,   lenglos2022multivariate, patow2022whole, trivedi2025interpretable}.

% #####################################################################
\subsection{Challenges and Open Problems}
% #####################################################################

As noted by \cite{alexandersen2026network}, several scientific open challenges demand attention in this context. These include questions on causal modeling of where and when pathology first emerges in a given individual's brain, whether the changes in neuronal activity observed alongside accumulating pathology compensatory or degenerative in nature, and can a unified model distinguish these possibilities from cross-sectional or short-longitudinal data alone. We also need to know how, and on what timescale, does neuronal stimulation  reduce protein burden and why do some individuals maintain cognitive function despite a high measured burden of pathology. 

Data science challenges involve joint estimation and uncertainty quantification involving high dimensional parameters  from comparatively sparse longitudinal neuroimaging data.  There are concerns about parameter identifiability: distinct parameter combinations may produce nearly indistinguishable predicted spatiotemporal trajectories, particularly when the true disease-onset time (which precedes the first available scan by an unknown and possibly substantial number of years) is itself unknown. The Bayesian approach of \cite{schafer2021} directly addresses this concern by propagating parameter uncertainty into predictive uncertainty, but the broader identifiability of the field's reaction-diffusion and multifactorial models, particularly under realistic clinical sampling schedules, remains incompletely characterized. Sparse and irregular longitudinal sampling presents a major statistical and ML challenge, as do multimodal data integration, model validation and computational scalability.

A  forward-looking direction that draws together many of the themes presented above is the development of \emph{virtual} or \emph{digital brain twins}: personalized, generative, and adaptive whole-brain models or one of its concrete Hopf, R\"ossler, or Wilson--Cowan instantiations  that are individually calibrated to a given subject's structural and functional imaging data and subsequently used for scientific and clinical inference \citep{zimmermann2022, wang2024virtualbraintwins}. Development of such digital twins remain a major challenge.

\section{Concluding comments}
\label{Sec:Conclusions}

This review has surveyed a broad and methodologically diverse toolkit for translational neuroscience and personalized neuro-health, spanning classical and Bayesian statistical regression, empirical-likelihood and covariate-assisted principal-component methods, geometric and topological network analysis, artificial intelligence, and dynamical-systems modeling. While these methodological traditions have largely developed in parallel --- within statistics, applied mathematics, computer science, and computational neuroscience, respectively --- the applications reviewed here illustrate their increasing convergence around a shared set of translational goals: earlier and more sensitive detection of neurodegenerative disease, individually meaningful (rather than purely population-averaged) characterizations of brain organization, and mechanistically interpretable, clinically actionable models of disease progression.

Several cross-cutting themes recur throughout this review. First, the movement from purely inferential, population-level, and stationary descriptions of brain activity toward personalized and dynamic characterizations is evident across every section: covariate-assisted principal regression and its Bayesian extension (Section~2.4) explicitly link functional connectivity to individual demographic and clinical covariates; curvature-based and topological descriptors (Section~3) are increasingly used as individual-level biomarkers rather than only group-level summaries; deep-learning and generative AI methods (Section~4) are applied directly to single-subject diagnosis, staging, and progression prediction; and dynamical-systems and digital-twin approaches (Section~5) aim explicitly at subject-specific, calibrated computational models of brain function. Second, geometric, topological, and dynamical-systems representations offer a valuable complement, rather than a replacement, to classical voxel- and region-wise statistical inference, often revealing structure --- network fragility, multiscale topological organization, or mechanistic disease trajectories --- that is not readily apparent from conventional graph-theoretical or mass-univariate analyses. Third, artificial intelligence and mechanistic dynamical modeling are themselves converging: generative models are increasingly used to simulate disease progression and to generate synthetic training data for mechanistic model calibration, while learned representations from deep architectures are being incorporated directly into dynamical-systems and digital-twin frameworks.

At the same time, this review has highlighted substantial open challenges that must be addressed before these methods can be reliably translated into clinical practice. Statistically, questions of model misspecification, multiple comparisons, and the reliability of thresholding and inference procedures remain active concerns even within the classical and Bayesian regression frameworks that anchor this literature \citep{eklund2016, friston2012ironic, lindquist2013ironic}. Methodologically, many of the geometric, topological, generative-AI, and mechanistic dynamical-systems approaches reviewed here require further work on identifiability, robustness to preprocessing and acquisition variability, and principled uncertainty quantification, particularly given the modest sample sizes and sparse, irregular longitudinal sampling that are typical of clinical neuroimaging studies. Computationally, scaling Bayesian, topological, and dynamical-systems methods to whole-brain, voxel-wise resolution and to large, multi-site cohorts remains demanding. And translationally, realizing the promise of personalized digital brain twins \citep{wang2024virtualbraintwins, zimmermann2022} will require not only continued methodological innovation but also principled frameworks for multimodal data integration, clinical and regulatory validation, and interoperable, secure data sharing across institutions, an issue directly addressed by the knowledge-graph and semantic-web approaches discussed in Section~3.3.

Taken together, the techniques reviewed here --- classical and Bayesian regression, empirical likelihood, covariate-assisted dimension reduction, network curvature and topological data analysis, knowledge graphs, artificial intelligence, and dynamical-systems and digital-twin modeling --- represent complementary and increasingly interconnected pieces of a common translational research agenda. Continued collaboration across statistics, applied mathematics, computer science, and clinical neuroscience will be essential to convert this rapidly growing methodological toolkit into reliable, interpretable, and clinically deployable tools for the early detection, monitoring, and personalized management of neurodegenerative disease.

\section*{Conflict of Interest Statement}

The authors declare no conflict of interest. 

\section*{Author Contributions}

VS, SN, SC and AC prepared Section~\ref{Sec:fMRI_Basics}, with SC leading the effort of Section~\ref{Sec:SC:EL} and SN leading the effort in Section~\ref{Sec:SN:CAP}. HF and CL led the work of Section~\ref{Sec:HF:NetworkCurvature}, SNKR and AD were the lead for Section~\ref{Sec:AD:TDA}, while SuC, AK and KJ spearheaded the Section~\ref{Sec:KJ:KG}. AC and VS led the work of Section~\ref{Sec:AI_Neuro}, with VS leading the effort for Section~\ref{Sec:VS:ChangeDetection}. For Section~\ref{Sec:DynSys_Neuro}, the primary lead was AB, AC, SN and PS. The coordination of the manuscript and the writing effort was led by AC. All authors contributed towards the writing and editing of this paper.

\section*{Funding}
Partial funding for this research is from the US National Science  Foundation grants DMS-2413491, DMS-2436549, DMS-2436550, DMS-2453756, DMS-2515815. 

\section*{Acknowledgments}
Authors acknowledge the use of large language model-based software products to search for some of the references, prepare bibliographic files and for some minor grammatical help with some parts of this manuscript.

\bibliographystyle{plain}
\bibliography{./Fuck}

\end{document}